\documentclass[11pt]{article}
\pdfoutput=1
\usepackage{jheppub}

\usepackage{slashed}
\usepackage{color, verbatim}
\usepackage{latexsym}
\usepackage{epsfig}
\usepackage{amsmath,accents}

\usepackage{amssymb}
\usepackage{graphicx}
\usepackage{bm}
\usepackage{stackrel}

\usepackage[font={small}]{caption}

\usepackage{hyperref}
\usepackage{epstopdf}
\definecolor{myred}{rgb}{0.7, 0, 0}
\definecolor{myblue}{rgb}{0, 0, 0.7}
\definecolor{mygreen}{rgb}{0.04, 0.7, 0.5}
\hypersetup{colorlinks,citecolor=myred,linkcolor=myblue,urlcolor=myblue,linktocpage=true}

\makeatletter

\@addtoreset{equation}{section}
\makeatother

\newcommand{\be}{\begin{equation}}
\newcommand{\ee}{\end{equation}}
\newcommand{\bea}{\begin{eqnarray}}
\newcommand{\eea}{\end{eqnarray}}

\newcommand{\cm}{{\textrm{cm}}}

\newcommand{\FI}{{\textrm{FI}}}
\newcommand{\FO}{{\textrm{FO}}}

\newcommand{\trho}{{\tilde\rho}}

\begin{document}

\thispagestyle{empty}

\begin{center}


\begin{center}

\vspace{.5cm}

{\Large\sc
Dark Branes for Dark Matter
\vspace{0.3cm}
}\\

\end{center}

\vspace{1.cm}

\textbf{
Fotis Koutroulis$^{\,a}$, Eugenio Meg\'{\i}as$^{\,b}$, Stefan Pokorski$^{\,a}$, Mariano Quir\'os$^{\,c}$
}\\

\vspace{1cm}

${}^a\!\!$ {\em {Institute of Theoretical Physics, Faculty of Physics,
University of Warsaw, Pasteura 5, PL 02-093, Warsaw, Poland}}

${}^b\!\!$ {\em {Departamento de F\'{\i}sica At\'omica, Molecular y Nuclear and Instituto Carlos I de F\'{\i}sica Te\'orica y Computacional, Universidad de Granada, Av. de Fuente Nueva s/n,  18071 Granada, Spain}}

${}^c\!\!$ {\em {Institut de F\'{\i}sica d'Altes Energies (IFAE) and The Barcelona Institute of  Science and Technology (BIST), Campus UAB, 08193 Bellaterra (Barcelona), Spain}}

\end{center}

\vspace{0.8cm}

\centerline{\bf Abstract}
\vspace{2 mm}

\begin{quote}\small
We propose a setup for the origin of dark matter based on spacetime with a warped extra dimension and three branes: the Planck brane, the TeV brane, at a (few) TeV scale $\rho_T$, and a dark brane, at a (sub)-GeV scale $\rho_1\lesssim 100$ GeV $\ll\rho_T$. The Standard Model is localized in the TeV brane, thus solving the Higgs hierarchy problem, while the dark matter $\chi$, a Dirac fermion with mass $m_\chi<\rho_1$, is localized in the dark brane. The radion, with mass $m_r<m_\chi$, interacts strongly ($\sim m_\chi/\rho_1\sim\mathcal O(1)$) with dark matter and very weakly ($\sim m_{f}\rho_1/\rho_T^2\ll 1$) with the Standard Model matter $f$.  The generic conflict between the bounds on its detection signatures and its proper relic abundance is avoided as dark matter annihilation is $p$-wave suppressed.  The former is determined by its very weak interactions with the SM and the latter by its much stronger annihilation into radions. Therefore, there is a vast range in the Dark Matter's parameter space where the correct relic abundance is achieved consistently with the existing bounds. Moreover, 
for the dark brane with $\rho_1\lesssim 3$~GeV,  a confinement/deconfinement first order phase transition, where the radion condensates,  produces a stochastic gravitational waves background at the nanoHz frequencies,  which can be identified with the signal detected by the Pulsar Timing Array (PTA) experiments. In the PTA window, for $0.15 \textrm{ GeV}\lesssim m_\chi\lesssim 2$ GeV the relic abundance is reproduced and all constraints are satisfied. 
\end{quote}

\vfill

\newpage

\tableofcontents

\newpage
\section{Introduction}
\label{sec:introduction}

The Standard Model (SM) of electroweak and strong interactions is known to be the theory of all non-gravitational interactions, as has been tested by a plethora of experimental results at high and low energy accelerators~\cite{ParticleDataGroup:2022pth}. Still, some theoretical (as \textit{e.g.}~the ultraviolet (UV) sensitivity, aka hierarchy problem, or the generation of the baryon asymmetry of the universe) and observational problems (mainly the existence of dark matter (DM) in the universe) lead us to believe that the SM is not the ultimate theory, but there should be some beyond the SM (BSM) physics. A very successful attempt to solve the hierarchy problem, \textit{i.e.}~understanding the hierarchy between the Planck and TeV scales, was done by Randall and Sundrum (RS)~\cite{Randall:1999ee} where a warped extra dimension with two branes (one at the Planck scale and another one at the TeV scale) is proposed such that the warp factor can relate both scales. In its simplest version~\cite{Randall:1999ee} the SM is living in the TeV brane (thus solving the Higgs hierarchy problem) while gravitons, as well as the radion fixing the brane distance~\cite{Goldberger:1999uk}, are propagating in the five-dimensional (5D) bulk. In fact the RS construction opens up the possibility of having an arbitrary number of branes~\cite{Lee:2021wau,Girmohanta:2023sjv}, such that the warp factor can ``explain" the natural mass scale for matter living in the brane. This setup is a perfect playground for hidden sectors, in cases where matter is localized on the TeV brane, as the former localized on different branes interacts with the SM only via gravitational interactions.

On the other hand the presence of DM has been confirmed by many observations at galactic and sub-galactic scales, including DM Milky Way and nearby galaxies and halos. In fact DM is observed in gravitationally collapsed structures of size ranging from the smallest known galaxies~\cite{Simon_2019} to galaxies of size comparable to the Milky Way~\cite{Salucci_2019}, to groups and clusters of galaxies~\cite{Allen_2011}. Candidates of DM are electrically neutral and colorless BSM particles, while our knowledge on DM requires that they have most likely gravitational interactions. There have been a plethora of candidates to DM, many of them, as WIMPs, getting their relic density by interactions with the SM which freeze-out for temperatures below the DM mass. A typical example of such WIMPs being the lightest neutralino in supersymmetric theories with R-parity~\cite{Jungman:1995df}. In those cases direct and indirect detection, as well as detection in accelerator searches are putting very severe bounds on the DM mass and annihilation cross-sections which jeopardize the corresponding theories, with a generic conflict between the correct relic abundance and those bounds. A way out for this problem is of course assuming that the DM is a SM singlet,  with only gravitational interactions to the SM. Once again RS-like theories offer an ideal framework for this class of theories as fields localized on different branes have only gravitational interactions, and thus possible interactions with SM fields, as \textit{e.g.}~the Higgs portal, can easily be suppressed by gravitational interactions. A DM model along these lines is proposed in this paper. The dominant channel responsible for the dark matter relic abundance is its annihilation into radions, whereas its detection signatures depend on its interactions with the SM.

Following the previous comments, we propose in this paper a Dark Sector (DS) localized in the dark brane (DB) at the infrared (IR) scale $\rho_1\lesssim ~100 \textrm{ GeV} \ll\rho_T$,  where $\rho_T$ is the scale of the SM brane.  In particular the dark brane contains the DM. As for the nature of the DM the most natural possibility is assuming a Dirac fermion $\chi$ with the requirement that its mass is $m_\chi\lesssim \rho_1$, as a Planckian tree-level mass is warped down to $\sim\rho_1$ in the four-dimensional theory. Smaller masses ($m_\chi\ll\rho_1$) are natural in the sense of 't Hooft~\cite{tHooft:1979rat}, as the symmetry of the theory increases in the limit $m_\chi\to 0$, and the mass parameter $m_\chi$ renormalizes multiplicatively~\footnote{Of course this property is not shared by the (perhaps) simpler possibility of DM being a real scalar $S$.}. 

We find that in the range $\rho_1\lesssim 100$ GeV the relic abundance $\Omega_\chi h^2=0.12$ can be obtained for  $m_\chi < {\cal O}(10)$ GeV.  It is worth noticing that, recently, the pulsar timing array (PTA) experiments detected a stochastic gravitational waves background~\cite{Xu:2023wog,Reardon:2023gzh,EPTA:2023fyk,NANOGrav:2023gor} which was proven to be consistent with that generated by the first order confinement/deconfinement phase transition triggered by a dark brane at the extreme IR, for scales in the interval  $\rho_1\in[10 \textrm{ MeV},3\textrm{ GeV}]$~\cite{Megias:2023kiy}.  In our model,  this range of $\rho_1$ is consistent with  the correct value of $\Omega_\chi h^2$ for $m_\chi < 2$ GeV. The existence of a DB in the range $10~ {\rm MeV} \lesssim \rho_1 \lesssim {\cal O} (10)~ {\rm GeV}$, with an intermediate brane at TeV scales, is then an extremely appealing possibility to accommodate a Dark Sector as this setup provides a natural and universal explanation for the hierarchy problem, the nature of Dark Matter and for the PTA measurements.

We will be agnostics about the origin of the mass term, as it can be simply an invariant Dirac mass, or can have its origin in some spontaneous symmetry breaking in the DS localized in the DB and some Yukawa coupling. For the computational sake we will just consider in this paper the case of a Dirac mass as $\mathcal L_{\rm DS}=-m_\chi \bar\chi\chi$, assuming in that way that only Dark Matter lives on the DB, but of course the dark sector can be more complicated, and be responsible of keeping the radion (strongly coupled to the dark brane) in thermal equilibrium.

In summary, in this paper we propose to study the dark matter capabilities of a setup permitting to generate the stochastic gravitational wave signal at nanoHz frequencies recently detected by the PTA collaborations. In a nutshell the main achievements and novelties of our model are: 
\begin{itemize}
\item
The dark matter only has direct gravitational interactions, with radions and massive KK gravitons. These interactions are strong enough that annihilations into radions can trigger the observed relic density after the non-relativistic freeze-out, while annihilations into gravitons are kinematically forbidden.  
\item
Gravitons interact with the SM much more weakly than radions which are the mediators between dark matter and the SM. Interactions of radions with the SM are weak enough to evade constraints from direct measurements, but not so weak as to also evade the neutrino floor, leaving a wide window for future experimental detection, mainly from nuclear recoil.
\item
The dark matter mass window, $0.15-2$ GeV, consistent with all direct and indirect constraints will allow to sharply concentrate the experimental searches.
\item
A spinoff is the prediction of a light radion which, in the future, can be detected in present fixed target experiments, as NA64 at the CERN SPS, and the future LDMX at SLAC.
\item
Finally, assuming that the PTA experiments have found a new physics scale around the GeV ($\Lambda_{\rm PTA}\sim$ GeV), our proposal would suggest that the new scale can be provided by the dark matter sector in our universe. 
\end{itemize}

As a last remark we would like to acknowledge that similar constructions, where different sectors are localized in different branes, interacting via 5D KK gravitons and radions, but which differ from ours, have been worked out in the past. Starting with two branes scenarios,  in Ref.~\cite{Lee:2013bua} the DM along the Higgs field are localized in the TeV brane, matter fields are in the Planck brane and gauge and gravity fields in the bulk. In this model the DM mass is linked to the electroweak symmetry breaking mechanism. In Ref.~\cite{Folgado:2019sgz} the SM and the (scalar) DM are in the TeV brane while only KK gravitons and radions propagate in the bulk. Dark matter is assumed in the 1-10 TeV range. A similar construction with two branes and with both the SM and (scalar and fermionic) DM in the IR brane is studied in Refs.~\cite{deGiorgi:2021xvm,deGiorgi:2022yha}. Dark branes have also been proposed at the GeV or sub-GeV scales in a number of papers. In Ref.~\cite{vonHarling:2012sz} the SM and DM are in the Planck brane (thus not solving the hierarchy problem) while in the bulk there is a $U(1)_X$, under which the DM is charged, and there is a dark brane at the GeV scale. Moving now to a three branes case, in Ref.~\cite{Ferrante:2023bcz} matter is in the Planck brane, the Higgs in the TeV brane to cope with the hierarchy problem, and the scalar $S$ DM is localized in the dark brane at the GeV scale. Note however that this three branes DM model differs from ours in many important aspects and in particular in the mechanism of the relic abundance production (see the discussion in Sec.~\ref{sec:conclusions} for more details).

The outline of the paper is as follows. In Sec.~\ref{sec:setup} we present the 5D setup, including the gravitational sector and the couplings of the graviton Kaluza-Klein (KK) and the radion modes to matter in the different branes. In Sec.~\ref{sec:abundance} we compute the DM annihilation into SM fields, which is subleading because the wave functions of KK gravitons and radions are very suppressed at the TeV brane, while the dominant DM annihilation channel is in a pair of radions. This annihilation channel keeps DM and radion in thermal equilibrium until the correct relic abundance is achieved. In Sec.~\ref{sec:direct} we confront the theory with direct detection experiments, for nuclear recoil experiments for $m_\chi\gtrsim 1$ GeV, and for electron recoil in atomic targets for $m_\chi\lesssim 1$ GeV. In both cases we find wide regions, in the plane $(m_\chi,m_r)$, where the model is allowed by experimental data, avoiding the neutrino floor region in the case of DM-nucleon experiments. Accelerator searches, in particular those based in events with missing energy and a $Z$ boson at the LHC and those based on fixed-target experiments, as NA64 at CERN, are considered in Sec.~\ref{sec:accelerator}. Indirect constraints are considered in Sec.~\ref{sec:indirect}, including those from Big Bang Nucleosynthesis, Planck constraints on cosmic microwave backgrounds and indirect measurements from cosmic rays. Finally our outlook and summary is given in Sec.~\ref{sec:conclusions}.

\section{The setup}
\label{sec:setup}
In this section we will describe the setup for the DM and the dark sector. We will consider a warped extra dimension~\cite{Randall:1999ee} with three branes~\cite{Cai:2021mrw,Lee:2021wau,Girmohanta:2023sjv,Megias:2023kiy}: the UV brane $\mathcal B_0$, at the scale $M_{\rm Pl}$, the TeV brane $\mathcal B_T$, at the TeV scale, and an extra IR brane $\mathcal B_1$, the dark brane, at some scale $\rho_1<$ TeV. The distance between the UV and TeV branes is fixed by the vacuum expectation value (VEV) of a (heavy) radion $R(x)$, which appears as an excitation of the 5D metric, with a mass $m_R\ll $ TeV, while the distance between the TeV and IR branes is fixed by the VEV of a (light) radion $r(x)$, with a mass $m_r\ll \rho_1$~\cite{Girmohanta:2023sjv}. The warp factor takes care of the hierarchy between the Planck and TeV scales, so that we will assume the Standard Model (SM) is  localized in the TeV brane, such that the Higgs hierarchy problem is solved. In a similar way the natural scale for possible scalars in the IR brane $\mathcal B_1$ would be fixed by the warp factor at the scale $\rho_1$, so the corresponding hierarchy problem, from the TeV scale to the scale $\rho_1$, would be equally solved by the theory warp factor. We will not consider other options as they will not be relevant for our work here.

We will now assume there is a dark sector (DS) which is  localized in the IR brane. The DS should contain a candidate to DM, $\chi$, with a mass $m_\chi$, which we are here assuming to be a Dirac fermion~\footnote{Other possibilities would be worth investigating, as \textit{e.g.}~the case of a scalar or a Majorana fermion.}. The fermion $\chi$ is supposed to be odd under some discrete symmetry, \textit{e.g.}~$\mathbb Z_2$, so that it is stable, while its relic abundance is fixed by its interactions with the radion $r(x)$ as we will see in the next section. Models of DM in DS's have already been considered, \textit{e.g.}~in Refs.~\cite{vonHarling:2012sz,Foot:2014uba,Breitbach:2018ddu,Fairbairn:2019xog,Brax:2019koq,Bento:2023flt}. As the SM is localized in the TeV brane and the DS in the IR brane we will just consider the 5D interval between both branes, and will not be concerned for the interval between the Planck and the TeV branes. 

As stated above the distance between the TeV and IR branes is fixed by the VEV of a light radion $r(x)$. The interaction between the IR and TeV branes is mediated by the gravitational sector propagating in the 5D bulk between both branes: the radion zero mode, $r(x)$, as well as the radion Kaluza-Klein (KK) modes $r_n(x)$, and the graviton zero mode $h_{\mu\nu}(x)$ and its KK excitations $h^n_{\mu\nu}(x)$. All of them will mediate the interaction between the SM and the DM $\chi$, and of course also all other possible fields in the DS.

The 5D theory between the TeV and IR branes is AdS$_5$ with some deformation, which is required to fix the brane distance and give a non-vanishing mass to the radion $r(x)$. It is usual to consider the theory determined by a superpotential $W(\phi)$~\cite{DeWolfe:1999cp}, where $\phi$ is the Goldberger-Wise stabilizing field~\cite{Goldberger:1999uk}, which transforms the second order Einstein equations in a system of first order ones. These theories have been extensively studied~\cite{Batell:2008me,Cabrer:2009we,Cabrer:2010si,Cabrer:2011fb,Cabrer:2011vu,Cabrer:2011qb,Megias:2020vek} so that here we will just give some elements which are necessary for the calculation of the couplings in this paper.

The 5D theory depends on the superpotential choice, as it does its predictions for the different spectra and couplings. Here we will consider the simplest example of a polynomial superpotential which reproduces the AdS$_5$ behavior in the UV, thus solving the hierarchy problem, while spoiling conformal invariance in the IR, thus allowing to fix the inter-brane distance and giving a mass to the corresponding radion.  We will use then the superpotential
\be
W(\phi)=6 k M_5^3+uk\phi^2 \,,
\ee
where $k$ is a mass dimension parameter which fixes the 5D curvature of the AdS space, $M_5$ is the 5D Planck scale, the 5D field $\phi$ has mass dimension $3/2$ and $u$ is a dimensionless parameter which creates the backreaction on the metric and which is responsible for the value of the radion mass, so it is usually assumed to be $u\ll 1$. Moreover the field $\phi$ is fixed to have values $v_\alpha$ by a potential at each brane $\mathcal B_\alpha$.

The solution to the 5D Einstein equations in conformal coordinates, \textit{i.e.}~such that the line element is
\be
ds^2=e^{-2A(z)}\eta_{MN}dx^M dx^N  \,,
\ee
where $z$ is the extra dimension in conformal coordinates, and $\eta_{MN}$ is the 5D flat metric. The metric 
in the region between the TeV and IR branes, is such that
\be
A(z)\simeq \log(kz),\quad \phi(z)=v_T (z/z_T)^u  \,,
\ee
so that, in the limit $u\to 0$, $\phi$ becomes a constant and the metric becomes AdS$_5$. The TeV and IR branes are located along the extra dimension at $z=z_T$ and $z=z_1$, respectively, and the corresponding branes scales are defined as $\rho_T\equiv 1/z_T$ and $\rho_1\equiv 1/z_1$.

\subsection{The graviton sector}
\noindent

The graviton appears from the metric tensor fluctuation $h_{\mu\nu}(z,x)$ that we will consider in the transverse traceless gauge: $\partial^\mu h_{\mu\nu}=h^\mu_\mu=0$.
The graviton coupling to matter fields localized in a  brane $\mathcal B_b$ is given by
\begin{equation}
\mathcal L=-\frac{1}{M_5^{3/2}}h_{\mu\nu}(z_b,x)T^{\mu\nu}_b(x) \,,
\end{equation}
where $T^{\mu\nu}_b(x)$ is the energy-momentum tensor of the matter localized in the $\mathcal B_b$ brane.
The equations of motion of the graviton sector can be found \textit{e.g.}~in Ref.~\cite{Davoudiasl:1999jd}. There is a massless zero mode, the physical graviton, while the mass spectrum of the KK modes is given by $m^h_n=x_n\rho_1$, where $x_n$ is determined by the condition $J_1(x_n)=0$, where $J_n(x)$ is a Bessel function of the first kind. As for the wave function of KK modes, after making the expansion in modes
\begin{equation}
h_{\mu\nu}(z,x) = \sum_{n=0}^\infty h^{(n)}(z) h^{(n)}_{\mu\nu}(x)  \,,
  \end{equation}
one easily finds that the (massless) graviton zero mode is constant $h^{(0)}(z)=\sqrt{k}$ while the KK modes are given by
\begin{equation}
h^{(n)}(z)=\sqrt{k}\frac{(kz)^2}{kz_1}\, \frac{J_2(m^h_n z)}{J_2(m^h_n z_1)} \,.
\end{equation}
 The coupling of all modes to the $\mathcal B_b$ brane is then written as
 \begin{equation} 
 \mathcal L=-\frac{1}{M_{\rm Pl}}h_{\mu\nu}^{(0)}(x)T^{\mu\nu}_{b}(x)-\sum_{n=1}^\infty c_n(z_b) h_{\mu\nu}^{(n)}(x) T^{\mu\nu}_b(x) \,,\qquad c_n(z_b)=\frac{1}{M_{\rm Pl}}\frac{(k z_b)^2}{kz_1}\frac{J_2(x_n z_b/z_1)}{J_2(x_n)} \,,
 \end{equation}
where we see that the coupling of the 4D graviton is suppressed as $1/M_{\rm Pl}$ so that its contribution to our present calculation can be safely neglected. 

The coupling then to the $\mathcal B_1$ brane is given, taking $z_b=z_1$, by~\footnote{Given the relation between $k$ and $N$ in the holographic theory, we have considered the range $1\lesssim k/M_P\lesssim 0.1$, such that $\rho_1\lesssim \tilde\rho_1\lesssim 10\rho_1$.}
\be
c_n(z_1)=\frac{kz_1}{M_{\rm Pl}}=\frac{k}{M_{\rm Pl}}\,\frac{1}{\rho_1}\equiv \frac{1}{\tilde\rho_1} \,
\label{eq:couplingB1}
\ee
(which implicitly provides the definition of the scale $\tilde\rho_b$ in terms of the brane scale $\rho_b$), 
so that it is stronger than the coupling of KK gravitons to the TeV brane in usual Randall-Sundrum models where the suppression is $\sim1/\rho_T$~\cite{Randall:1999ee}, with $\rho_T\simeq $ few TeV. As for the coupling to the $\mathcal B_T$ brane,
assuming that $z_T/z_1\ll 1$ we can approximate the coupling $c_n(z_T)$ as
\be
c_n(z_T)\simeq \frac{x_n^2}{8 J_2(x_n)}\frac{1}{\tilde\rho_T}\left( \frac{\tilde\rho_1}{\tilde\rho_T}\right)^3 \,,\quad \textrm{and}\quad 
c_n(z_T)c_n(z_1)=\frac{x_n^2}{8 J_2(x_n)}\frac{\tilde\rho_1^2}{\tilde\rho_T^4}  \,.
\ee
In particular for the first KK mode we have $x_1=3.83$ so that 
\be
c_1(z_T)\simeq 4.6 \frac{1}{\tilde\rho_T}\left( \frac{\tilde\rho_1}{\tilde\rho_T}\right)^3,\quad \textrm{and}\quad c_1(z_1) c_1(z_T)\simeq 4.6 \frac{1}{\tilde\rho_T^2} \left(\frac{\tilde\rho_1}{\tilde\rho_T} \right)^2 \,.
\label{eq:c1cTgrav}
\ee

For computing the interesting cross-sections, the KK gravitons with momentum $q$ will propagate between the $\mathcal B_T$ and $\mathcal B_1$ branes with a propagator~\cite{Contino:2001nj}
\be
P_{\mu\nu;\rho\sigma}^{\,(n)}(q)=\frac{i}{q^2-(m_n^h)^2}\left(\frac{1}{2}t_{\mu\rho}t_{\nu\sigma}+\frac{1}{2}t_{\mu\sigma}t_{\nu\rho}-\frac{1}{3}t_{\mu\nu}t_{\rho\sigma}  \right),\quad t_{\mu\nu}=\eta_{\mu\nu}-\frac{q_\mu q_\nu}{(m_n^{h})^2}  \,,
\ee
connecting the DM in the brane $\mathcal B_1$ with ordinary matter in the brane $\mathcal B_T$. The propagator satisfies the traceless and transverse conditions
$
[\eta^{\mu\nu},\eta^{\rho\sigma},q^\mu,q^\nu,q^\rho,q^\sigma]P_{\mu\nu;\rho\sigma}(q)=0
$.

As the SM is in the $\mathcal B_T$ brane, with energy-momentum tensor $T^{\mu\nu}_{\rm SM}$, and the dark sector in the $\mathcal B_1$ brane, with energy-momentum tensor $T_{\rm DS}^{\mu\nu}$, the exchange of KK gravitons between both branes, with momentum transfer $q^2\ll m_n^2$, generates the effective Lagrangian given by 
\be
\mathcal L_{\rm eff}=\sum_{n=1}^\infty a_n \left(T_{\mu\nu}^{\rm SM}T^{\mu\nu}_{\rm DS} -\frac{1}{3}T_{\rm SM}T_{\rm DS} \right),\quad 
\textrm{where}\quad T=\eta_{\mu\nu}T^{\mu\nu},\quad a_n=\frac{c_n(z_T)c_n(z_1)}{(m_n^h)^2}  \,,
\label{eq:Leffgrav}
\ee
with, generically, dimension 8 operators.
The coefficient $a_{n}$ is very suppressed. In particular for the first mode $n=1$, it turns out to be
\be
a_{1}=0.32 \,(k/M_{\rm Pl})^{-2}\, \frac{1}{\tilde\rho_T^4} \,.
\ee

\subsection{The radion sector}

\noindent The radion corresponds to scalar perturbations $F(z,x)$ of the metric as $$ds^2=  e^{-2 A(z)} \left[ e^{-2F(z,x)} \eta_{\mu\nu} dx^\mu dx^\nu+(1 + 2 F(z,x))^2 dz^2\right].$$
The wave function of the KK modes of the light radion can be decomposed as
\begin{equation}
F(z,x) = \sum_{n=0}^\infty f^{(n)}(z)r^{(n)}(x) \,.
  \end{equation}

In this subsection we will consider the dynamics of the lightest radion $r^{(0)}(x)\equiv r(x)$ with a 5D profile $f^{(0)}(z)\equiv f(z)$. This mode is massless if we neglect the backreaction on the metrics. However, after the backreaction is taken into account, it can get a mass $m_r^2\ll m_n^2$. The value of the radion mass $m_r$ depends on the superpotential parameter $u$, as well as the details of the $\mathcal B_1$ localized potential (in particular on its second derivative) fixing the value of $\phi$ at the particular value $v_1$. A computation of the radion mass in the stiff wall limit leads to~\cite{Csaki:2000zn}
\begin{equation}
\frac{m_r}{\rho_1} = \frac{2}{\sqrt{3}} \bar v_1 u \,.  \label{eq:mr}
  \end{equation}
We have introduced the dimensionless quantity $\bar v_\alpha \equiv v_\alpha / M_5^{3/2}$, where $\alpha$ refers to the $\mathcal B_\alpha$ brane. Therefore we can consider the radion mass as a free parameter. 

The coupling of the radion zero mode to matter localized on the $\mathcal B_b$ brane is given by
\begin{equation}
\mathcal L= - c_r(z_b) r(x) T_{b}(x) \,,
\end{equation}
where $T_b(x)=\eta_{\mu\nu}T_b^{\mu\nu}$. 
In the limit of small backreaction, after imposing that the field $r(x)$ is canonically normalized, one gets that the function $c_r(z_b)$ is given by~\cite{Csaki:2007ns}
\begin{equation}
c_r(z_b)=\left(\frac{k}{M_{\rm Pl}} \right)\,\frac{1}{\sqrt{6}}\frac{z_b^2}{z_1} \,,
\end{equation}
so that the couplings to the $\mathcal B_T$ and $\mathcal B_1$ branes are given by
\be
c_r(z_T)= \frac{\tilde\rho_1}{\sqrt{6}\tilde\rho_T^2}  ,\quad c_r(z_1)=\frac{1}{\sqrt{6}\tilde\rho_1}\,,\quad \textrm{and}\quad c_r(z_T)c_r(z_1)=\frac{1}{6\tilde\rho_T^2}  \,.
\label{eq:c1cTrad}
\ee
Notice that for $\rho_1\ll \rho_T$, the product of couplings $c_r(z_T)c_r(z_1)$ is much larger than the corresponding product for graviton exchange in Eq.~(\ref{eq:c1cTgrav}). In particular the coupling to the SM brane $\mathcal B_T$ is suppressed with respect to the Randall-Sundrum case by the factor $\rho_1/\rho_T$, while the coupling to the DS brane is enhanced by the same factor.

As in the previous section, the exchange of a radion between the SM and DS branes generates the effective Lagrangian
\be
\mathcal L_{\rm eff}=a_r T_{\rm SM}T_{\rm DS} \,,\quad \textrm{where}\quad a_r= - \frac{c_r(z_T)c_r(z_1)}{q^2 - m_r^2} \, ,
\label{eq:Leffrad}
\ee
that will be used in Sec.~\ref{sec:direct}.
Note that for the radion exchange case we have kept the associated propagator in the general expression of the effective coupling, while we are neglecting the radion width $\Gamma_r$ which will be shown to be much smaller than the radion mass $\Gamma_r/m_r\ll 1$.
However, in the following when it comes to $a_r$ we will use either the small or the large momentum transfer, $q^2\ll m_r^2$ or $q^2\gg m_r^2$ respectively.

\section{The relic abundance}
\label{sec:abundance}

 In a previous paper, see Ref.~\cite{Megias:2023kiy}, we have seen that in the presence of the $\mathcal B_1$ brane, at the scale $\rho_1$, there is a confinement/deconfinement first order phase transition at the nucleation temperature $T_n\lesssim \rho_1$, followed 
by a reheating at a temperature $T_{\rm RH}\gtrsim \rho_1$. In the language of the AdS/CFT correspondence the radion and the dark sector (including the dark matter $\chi$) localized in the IR brane appear as composite states of a holographic conformal theory~\cite{Konstandin:2010cd,Megias:2020vek,Megias:2023kiy}. It was proven that the first order phase transition triggers a stochastic gravitational waves background which can be fitted with the recent observations from the Pulsar Timing Arrays (PTA) collaborations~\cite{NANOGrav:2023gor,EPTA:2023fyk,Reardon:2023gzh,Xu:2023wog} with nanoHz frequencies, provided that $\rho_1\in[10 \textrm{ MeV},3 \textrm{ GeV}]$, which translates into the bound $\tilde\rho_1\lesssim 30$ GeV.

In this section we investigate the relic abundance of the fermion $\chi$ under the assumption that it freezes out from the thermal bath consisting of the SM fields and the radion.  In Sec.~\ref{sec:indirect} we verify this assumption by solving the coupled Boltzmann equations for the $\chi$ and the radion.

 For the case of a thermal relic $\chi$ its energy density today $\Omega_\chi$ depends on its annihilation rate $\langle \sigma v\rangle$ as (see \textit{e.g.}~Ref.~\cite{Hooper:2018kfv}),
 \be
 \Omega_{\rm \chi}h^2\simeq 0.1\, \frac{x_{\FO}}{10}\sqrt{\frac{65}{g_\ast(T_{\FO})}} \frac{\langle \sigma v\rangle_c}{\langle \sigma v\rangle}
 ,\quad \textrm{with}\quad \langle\sigma v\rangle_c \sim 1.09 \times 10^{-9} \textrm{ GeV}^{-2}\,,
 \label{eq:DMc}
 \ee
where $x_{\FO}=m_\chi/T_{\FO}\gg 1$ (typical of a cold thermal relic) is provided by the freeze-out temperature such that $\langle\sigma v\rangle n_\chi(T_{\FO})\simeq H(T_{\FO})$, $H(T)$ is the Hubble parameter and $g_\ast(T)$ is the effective number of relativistic degrees of freedom at the temperature $T$.

The thermal average $\langle \sigma v\rangle$ is defined as
 \be
 \langle\sigma v\rangle=\frac{1}{8m_\chi^4 TK_2^2(m_\chi/T)}\int_{4m_\chi^2}^\infty (s-4m_\chi^2)\sqrt{s}K_1(\sqrt{s}/T)\sigma(s) ds  \,,
 \label{eq:sigmav}
 \ee
 where $v$ is the M\o ller velocity~\footnote{The M\o ller velocity $v$ (sometimes referred to as $v_{\textrm{M\o l}}$) is related to the velocities of incoming particles $\vec v_1$ and $\vec v_2$ by $v=(|\vec v_1-\vec v_2|^2-|\vec v_1\times \vec v_2|^2)^{1/2}$~\cite{Gondolo:1990dk}, where $\vec v_i\equiv \vec p_i/E_i$. In the center of mass system $v=2|\vec v_1|$.}, and Eq.~(\ref{eq:sigmav}) is valid for $T\lesssim 3m_\chi$~\cite{Gondolo:1990dk}.

\subsection{Annihilation into SM fields}

Both the radion and the KK gravitons connect the DM, localized in $\mathcal B_1$, with the SM localized in the $\mathcal B_T$ brane, so in principle both (or in particular the radion and the first --the lightest-- KK mode graviton) can mediate the annihilation of DM into SM fields. Even if both couplings to the $\mathcal B_1$ brane are equal, $c_r(z_1)=c_1(z_1)=1/\trho_1$, the coupling of the graviton to the $\mathcal B_T$ brane is much smaller that that of the radion, $c_1(z_T)/c_r(z_T)\simeq (3 \trho_1/\trho_T)^2\ll 1$ for $\trho_1\ll \trho_T$. Moreover, given that we are using radion masses much smaller than the DM mass $m_r\ll m_\chi$, the ratio of annihilation cross sections into the SM mediated by KK gravitons is further suppressed with respect to that mediated by the radion by a factor $\sim (2 m_\chi/m_1^h)^4\simeq 7\times 10^{-2}(m_\chi/\rho_1)^4\ll 1$, as we are considering $m_\chi< \rho_1$. We will then consider the radion as the messenger between the DM and the SM.

As the radion couples to the SM through the trace of the energy momentum tensor, the first candidates for DM annihilation into SM fields are fermions $f$ with mass $m_f\lesssim m_\chi/2$ and massless gauge bosons, the gluon and the photon. In fact the coupling between $\chi$ in the $\mathcal B_1$ brane and the radion, $g_{r\chi\bar\chi}$, and the coupling between the SM fermion $f$ in the $\mathcal B_T$ brane and the radion, $g_{r f\bar f}$, are such that
 \be
 g_{r\chi\bar\chi}g_{r f \bar f}\simeq\frac{m_\chi m_f}{6\tilde\rho_T^2}  \,.
 \ee
The total cross-section for the process $\chi\bar\chi\to f\bar f$ mediated by the radion, is given by
\be
\sigma_{f}=( g_{r\chi\bar\chi}g_{r f \bar f})^2 \frac{1}{16\pi s}\left(1-\frac{4m_\chi^2}{s} \right)^{1/2}\left(1-\frac{4m_f^2}{s} \right)^{3/2}  \,,
\label{eq:sigmaf}
\ee
where we have neglected the radion mass, as $s\gg m_r^2$. Using Eq.~(\ref{eq:sigmaf}) and $s>4m_\chi^2$, $m_f<m_\chi/2$, we find an upper bound for $\sigma_f$ as
\be
\sigma_f \lesssim 10^{-4}\, \frac{m_\chi^2} {\tilde\rho_T^4}\,.
\ee
 Using now the considered range $\tilde\rho_T\gtrsim 1$ TeV and e.g.~$m_\chi\lesssim 2$ GeV, and $v<1$, we get the upper bound $\sigma_f v\lesssim 4\times 10^{-16} \textrm{ GeV}^{-2}$, much smaller than the value of $\sigma_c v$ in Eq.~(\ref{eq:DMc}).

Moreover, the radion is also coupled to the gluon with an effective vertex given by~\cite{Giudice:2000av}
\be
\mathcal L=c_r(z_T) \left(b_3-\frac{1}{2}\sum_{q}F_{1/2}(\tau_q) \right)\frac{\alpha_3}{8\pi}\, r(x)G_{\mu\nu}G^{\mu\nu}   \,,
\label{eq:effgg}
\ee
where $\tau_q=4 m_q^2/s$, $F_{1/2}(\tau_q)$ is defined in Ref.~\cite{Giudice:2000av} and has the property that $F_{1/2}(\tau_q)\simeq -4/3$, for $\tau_q\gg 1$, and $F_{1/2}(\tau_q)\simeq 0$ for $\tau_q\ll 1$. We are considering in Eq.~(\ref{eq:effgg}) the contribution of quarks that can be integrated out at the scale $\mathcal Q \sim m_\chi$, while we are neglecting the contribution from lighter quarks. The first term, where $b_3=7$ is the QCD $\beta$-function in the SM, comes from the QCD trace anomaly, while the second term comes from a loop diagram with virtual quarks. Notice that the coefficient $b_3-\frac{1}{2}\sum_{q}F_{1/2}(\tau_q)$, in the limit $\tau_q\to\infty$, is the QCD $\beta$-function in the presence of $6-n_f$ flavors, after decoupling $n_f$ flavors of quarks with masses $m_q>m_\chi$, $b_{\rm QCD}\simeq 7+2n_f/3$. Hereafter, we will consider the running of $\alpha_3(s)$ at the scale $s\simeq 4 m_\chi^2$.

Using the effective Lagrangian (\ref{eq:effgg}), the cross-section for the process $\chi\bar\chi\to gg$ is given by
\be
\sigma_g =\frac{1}{32\pi}\left[\frac{m_\chi}{\sqrt{6}\tilde\rho_T^2}\left(b_3-\frac{1}{2}\sum_{q}F_{1/2}(\tau_q) \right) \frac{\alpha_3}{8\pi} \right]^2 \left(1-\frac{4m_\chi^2}{s} \right)^{1/2}   \,,
\ee
which is also bounded, for $s>4m_\chi^2$ by
\be
\sigma_g < \frac{m_\chi^2}{192\pi \tilde\rho_T^4}\left( \frac{b_{\rm QCD}\,\alpha_3}{8\pi} \right)^2\simeq 2\times 10^{-4}\frac{m_\chi^2}{\tilde\rho_T^4}  \,,
\ee
where we have used $\alpha_3\lesssim 1$, so that the condition $\sigma_g v\ll \sigma_c v$ is always fulfilled.

The radion is also coupled to photons with an effective vertex given, in the limit $m_f^2/s\to\infty$, by~\cite{Giudice:2000av}
\be
\mathcal L=c_r(z_T) b_{\rm QED}\frac{\alpha_{\rm EM}}{8\pi}\, r(x)F_{\mu\nu}F^{\mu\nu} \,,
\ee
where  $b_{\rm QED}=-32/3+16 n_u/9+4 n_d/9+4 n_\ell/3$ is the QED $\beta$-function after decoupling $n_u$ flavors of up quarks, $n_d$ flavors of down quarks and $n_\ell$ flavors of leptons, with masses $m_f>m_\chi$. A similar calculation for the process $\chi\bar\chi\to\gamma\gamma$ gives for the cross section $\sigma_\gamma\simeq 10^{-4}\sigma_g$.

\begin{figure}[t]
  \centering
  \includegraphics[width=7.5cm]{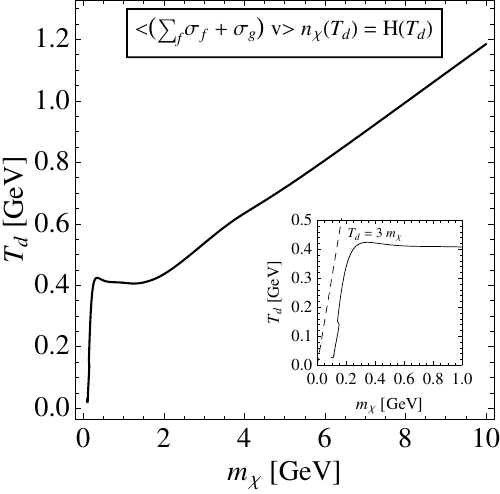} \hspace{0.5cm}
  \includegraphics[width=9.3cm]{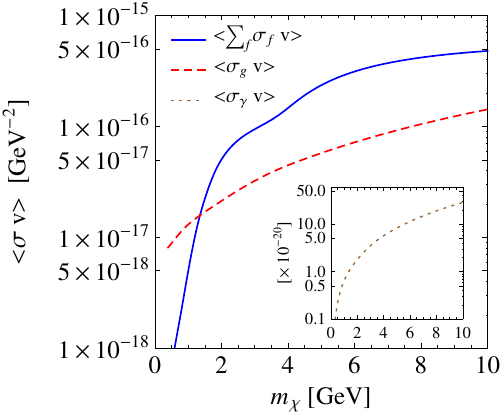} 
\caption{\it Left panel: Decoupling temperature $T_d$ of $\chi$ from the SM as a function of $m_\chi$. Right panel: Thermal average of cross-sections at the decoupling temperature  as a function of $m_\chi$. We have considered the channels $\chi\bar\chi\to f\bar f$, $\chi\bar\chi\to gg$ and $\chi\bar\chi\to \gamma\gamma$, as functions of $m_\chi$. For every value of $m_\chi$ the cross section $\sum_f\sigma_f$ is dominated by the heaviest fermion such that $m_f<m_\chi$.
We have used $\tilde\rho_T=1$~TeV.}
\label{fig:Td}
\end{figure}

As DM annihilation cross-sections into SM fields are dominated by the $f\bar f$, such that $m_f\lesssim m_\chi/2$, and $gg$ channels, we can then evaluate the temperature at which $\chi$ decouples from the SM. We will define the interaction rate as $\Gamma_\chi= \Big\langle \sum_f\sigma_f v+\sigma_g v \Big\rangle n_\chi$, where the number density for a gas of fermions is~\footnote{Notice that for a gas of bosons the number density is given by Eq.~(\ref{eq:nchi}) with $\xi = 1$.}
  \begin{equation}
    n_\chi(T) = \frac{1}{2\pi^2} g_\chi
     m_\chi^2 T \sum_{n=1}^\infty \frac{\xi^{n+1}}{n} K_2\left(n m_\chi/T\right)  \,, \label{eq:nchi}
  \end{equation}
  with $\xi = -1$, while $K_n(z)$ is the modified Bessel function of the second kind. This formula has the following relativistic $(m_\chi \ll T)$ and non-relativistic $(m_\chi \gg T)$ limits:
  \begin{equation}
    n_\chi(T) \simeq \left\{
    \begin{array}{cc}
    \frac{3}{4}\frac{\zeta(3)}{\pi^2} g_\chi T^3    &  \qquad m_\chi \ll T \\
    g_\chi \left(\frac{m_\chi T}{2\pi} \right)^{3/2} e^{-m_\chi/T}  & \qquad m_\chi \gg T 
    \end{array} \,.  \right.
  \end{equation}
  We have considered the general expression, Eq.~(\ref{eq:nchi}), in the computation of the decoupling temperature. The decoupling temperature $T_d$ is then obtained from the equation
  \be
  \Big\langle \sum_f \sigma_f v+\sigma_g v \Big\rangle n_\chi(T_d)\simeq H(T_d)=\pi\sqrt{\frac{g_\ast(T_d)}{90}}\,\frac{T^2_d}{M_{\rm Pl}} \,.
  \ee
The decoupling temperature $T_d$, for which $\Gamma_\chi\simeq H$, is exhibited in the left panel of Fig.~\ref{fig:Td}.

We can see that for $m_\chi<10$ GeV, the DM decouples from the SM at temperatures $T_d<1.2$ GeV, well in the non-relativistic regime. In the right panel of Fig.~\ref{fig:Td} we show the dependence of the thermal averaged cross section at the decoupling temperature $T_d$. We can see there that its typical values are such that  $\langle\sigma v\rangle\ll \langle\sigma v\rangle_c$.
From the previous analysis we can deduce that, when $\chi$ goes out of equilibrium with the SM its annihilation cross section is too small to generate the observed cosmic DM density observed today. The DM would then overclose the universe unless there is another more efficient annihilation channel. As we will see now this happens with the $rr$ channel.

\subsection{Annihilation into radions}
\label{sec:annihilation}

On top of the arguments presented in the previous section, here we also assume that the DM mass is $m_\chi \ll m^h_n$ suggesting that DM cannot annihilate non-relativistically into KK gravitons. Therefore, given the interaction between $\chi$ and the radion field $r$, there is an additional channel where the DM annihilates into a couple of radions, $\chi\bar\chi\to rr$, with the exchange of the fermion $\chi$ in the $t$- and $u$-channels, as well as the contact interaction term $\bar\chi\chi rr$, and we are here assuming that radions are in thermal equilibrium with the SM plasma. Unlike the previous channels into SM fields, the global square coupling is sensitive to the value of $\rho_1$. The Feynman rules for the relevant couplings in the process $\chi(p)+\bar\chi(p')\to r(k)+r(k')$, where $p,p'$ ($k,k'$) are incoming (outcoming) momenta, are~\cite{Folgado:2019sgz}
\begin{align}
\chi(p)\bar\chi(q)r(k) &  \quad \Rightarrow \quad -i\frac{8m_\chi-3(\slashed{p}+\slashed{q})}{2\sqrt{6}\tilde\rho_1} \equiv  -i\frac{5m_\chi-3\slashed{q}}{2\sqrt{6}\tilde\rho_1}\nonumber  \,, \\
\chi(q)\bar\chi(p')r(k') & \quad  \Rightarrow \quad -i\frac{8m_\chi-3(\slashed{q}-\slashed{p'})}{2\sqrt{6}\tilde\rho_1}\equiv-i\frac{5m_\chi-3\slashed{q}}{2\sqrt{6}\tilde\rho_1}  \,, \\
\bar\chi(p)\chi(p')r(k)r(k') & \quad \Rightarrow \quad  +i\frac{m_\chi}{12\tilde\rho_1^2}\nonumber  \,,
\end{align}
where $q=p-k$ and the Dirac equation has been used for on-shell fermions $\slashed{p}u(p)=m_\chi u(p)$ and anti-fermions $\bar v(p')\slashed{p}^\prime = -m_\chi\bar v(p^\prime)$.

The total cross section is given by
\be
\sigma_r = \frac{1}{1152\pi} \frac{m_\chi^2}{\tilde\rho_1^4}\left[\frac{z^2 (7 - 11 z^2 - z^4)}{(1-z^2)} \tanh^{-1}(\sqrt{1-z^2}) + \frac{169 - 121 z^2 - 8 z^4}{8 (1-z^2)^{1/2}}\right] \quad \textrm{with}\quad z^2=\frac{4m_\chi^2}{s}  \,,  \label{eq:sigmar}
\ee
where we are neglecting the radion mass $m_r$.~\footnote{Note that since the two radions in the final state are identical we integrate $\theta$ in the interval $0\leq \theta\leq \pi/2$.}

\begin{figure}[t]
\centering
\includegraphics[width=7.5cm]{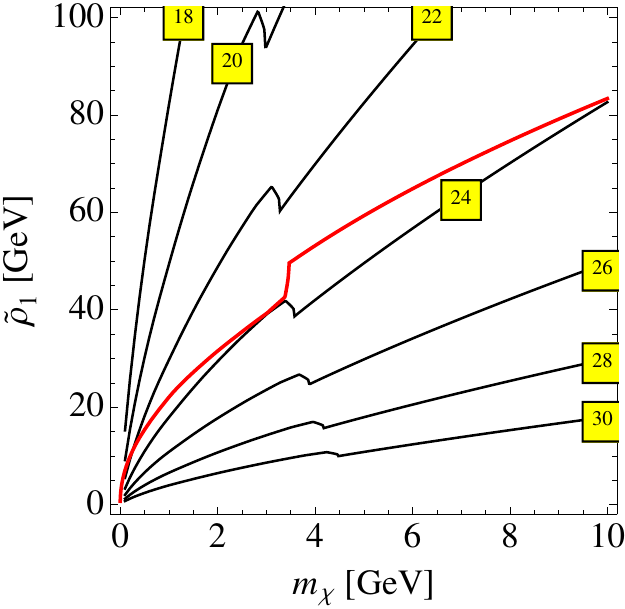}   \hspace{1.5cm} \includegraphics[width=7.8cm]{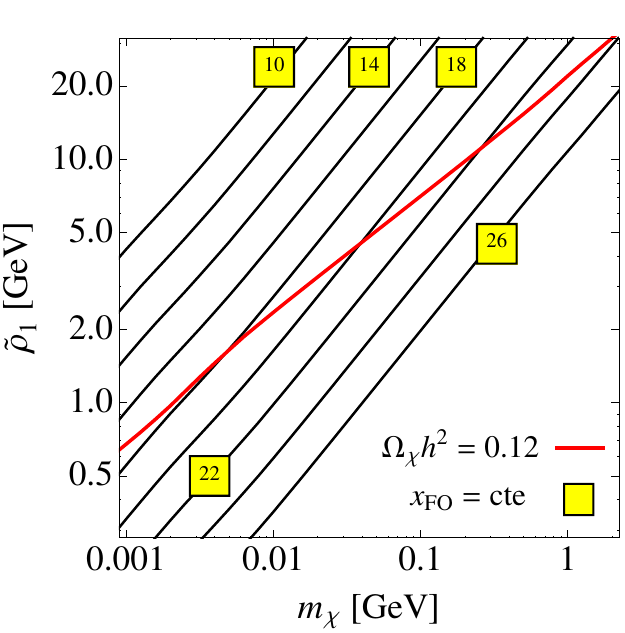}
\caption{\it Left panel: Contour levels of the freeze-out parameter $x_{\FO}=m_\chi/T_{\FO}$ (black solid lines) and the cosmic density $\Omega_\chi h^2=0.12$ (red solid line) for the radion production $\chi\bar\chi\to rr$ in the plane $(m_\chi,\trho_1)$. Right panel: Detail of the left panel in logarithmic coordinates, for the region where the confinement/deconfinement phase transition leading to the PTA signal takes place.}
\label{fig:Tr}
\end{figure} 
In the left  panel of Fig.~\ref{fig:Tr} we display, in the plane $(m_\chi,\tilde\rho_1)$, contour lines of the freeze-out parameter $x_{\FO}=m_\chi/T_{\FO}$ corresponding to the radion production $\chi\bar\chi\to rr$, where we can check that the freeze-out happens when the $\chi$ fermion is non-relativistic, as $x_{\FO}\gg 1$. The prediction for $\Omega_\chi h^2=0.12$ is provided by the red solid line. In general, we see that the correct relic abundance is obtained for $m_\chi\ll\tilde\rho_1$.  The origin of the laggedness of contour lines is the `jump' in the number of degrees of freedom around the QCD phase transition. In particular,  the region where we can explain the PTA data with the confinement/deconfinement phase transition, for $\rho_1\lesssim \mathcal  O(10)$ GeV~\cite{Megias:2023kiy}, is shown in the right panel of Fig.~\ref{fig:Tr}, where we get the correct cosmic density for very light $\chi$ masses, $m_\chi\lesssim 2$ GeV, with a freeze-out temperature $x_{\FO}\simeq \mathcal O(20)$, where the DM is non-relativistic. 

Some analytical expressions can be obtained by considering an expansion of $\sigma_r$ for $s \simeq 4 m_\chi^2$,
\begin{equation}
  \sigma_r \simeq \frac{803}{55296 \pi} \frac{m_\chi \sqrt{s}}{\tilde\rho_1^4} \sqrt{1 - z^2} \,,
\end{equation}
which means $\chi\bar\chi$ annihilation is in $p$-waves, together with a non-relativistic expansion of $\langle \sigma v\rangle_{\FO}$, \textit{i.e.}~for $x_{\FO} \gg 1$. One finds
\begin{equation}
x_{\FO}(m_\chi,\tilde\rho_1) \simeq \frac{1}{2} \mathcal W_0\left[ \frac{5}{2} \left( \frac{803}{12} \right)^2 \pi \frac{g^2_\chi}{g_\ast} \frac{M_{\rm Pl}^2 m_\chi^6}{(4\pi\tilde \rho_1)^8} \right] \,,
\end{equation}
while the dependence of $\tilde\rho_1$ with $m_\chi$ (at constant $x_{\FO}$ or $\Omega_\chi h^2$) are
\begin{eqnarray}
  \tilde\rho_1(m_\chi,x_{\FO}) &\simeq& \frac{1}{4} \left[ \frac{803 \sqrt{5}}{24 \pi^{7/2}} \frac{e^{-x_{\FO}}}{\sqrt{x_{\FO}}}  \frac{g_\chi}{\sqrt{g_\ast}} M_{\rm Pl} m_\chi^3 \right]^{1/4}  \,, \\
  \tilde\rho_1(m_\chi,\Omega_\chi h^2) &\simeq& \frac{1}{6} \left[ \frac{803}{16\pi} \frac{\sqrt{g_\ast} m_\chi^2 \, \Omega_\chi h^2}{c \langle \sigma v\rangle_c } \mathcal W_{-1}^{-2}\left( -\frac{4\pi}{3} \left( \frac{ \pi g_\ast \, \Omega_\chi h^2}{ \sqrt{90} \, c g_\chi M_{\rm Pl}  \langle \sigma v\rangle_c m_\chi}  \right)^{2/3} \right)  \right]^{1/4}  \,,
\end{eqnarray}
respectively. In these formulas $c = 0.1 \sqrt{65}/10 \simeq 0.08062$, and $\mathcal W_n(z)$ is the $n$-th branch of the Lambert function. The behavior is approximately $\tilde\rho_1(m_\chi,x_{\FO}) \propto m_\chi^{3/4}$ and $\tilde\rho_1(m_\chi,\Omega_\chi h^2) \propto m_\chi^{1/2}$, as it can be seen in Fig.~\ref{fig:Tr}, where the numerical results using the full expression of $\sigma_r$ given by Eq.~(\ref{eq:sigmar}) are shown.

Notice that our DM candidate, the $\chi$ fermion, looks like a WIMP particle, but much lighter and much more weakly coupled than typical WIMPs. In fact, while the DM has a typical scale around the GeV scale the interactions triggering thermal equilibrium with the radion are much weaker than those for heavier WIMPs. This possibility has been studied in full generality in Ref.~\cite{Feng:2008ya} where the corresponding DM particle was named WIMPless (and the mechanism dubbed ``WIMPless miracle"), and a supersymmetric model with gauge mediated supersymmetry breaking was proposed as a possible example. In our case the weakness of the interactions between DM and the radion is generated by a small hierarchy between the parameters $m_\chi$ and $\tilde\rho_1$. While the warp factor, responsible for generating the TeV scale at the brane $\mathcal B_T$ where the SM is living, will naturally generate, for matter living in the $\mathcal B_1$ brane, the scale $\rho_1$, values of $m_\chi\ll \tilde\rho_1$ are also natural~\footnote{This would not be the case for a scalar particle living in the IR brane.} as the fermion mass renormalizes multiplicatively and the radiatively corrected mass is proportional to the original tree-level one.

Two comments now about the annihilation cross-section of DM $\bar\chi\chi\to rr$:
\begin{itemize}
\item 
One could think about the possible annihilation channel into two KK gravitons as $\bar\chi\chi\to G_n G_n$. However this channel is closed for the values of the DM mass $m_\chi<m^h_1\simeq 3.8 \rho_1$ that we are considering in Fig.~\ref{fig:Tr}. 
  
\item We have considered the case of a light radion, which is consistent with the validity of the effective theory where all KK modes are integrated out. In particular we are considering $m_r\ll m_\chi$ which takes into account normal annihilation of $\bar\chi\chi\to rr$, and thus excludes the case of forbidden DM~\cite{DAgnolo:2015ujb} for which $m_r\sim m_\chi$, and which requires a certain amount of mass tuning. The line $m_r=m_\chi$ is shown in Fig.~\ref{fig:radion_bounds} where we can see it lies near the neutrino floor or in the forbidden region from DM direct detection in DM-nucleon cross-sections.
\end{itemize}

In the next section we will analyze how direct detection experiments can be in agreement with our theory. As we will see the radion mass $m_r$ which was irrelevant for the annihilation processes, as $s\gg m_r^2$, will be relevant for the direct detection processes which take place at a momentum transfer $q$ such that $q^2\ll m_r^2$.

\section{Direct detection}
\label{sec:direct}

As we have seen in previous sections after integrating out the graviton KK modes and the radion zero mode we get an effective Lagrangian which depends on the energy-momentum tensors $T^{\mu\nu}_{\rm SM}$ and $T^{\mu\nu}_{\rm DS}$. Regarding the possibility for direct DM detection based on its scattering off nuclei, we will be concerned on quarks $Q$ localized on the $\mathcal B_T$ brane.  In Sec.~\ref{sec:abundance} we have focused on the range $\tilde\rho_1<100$ GeV (as we are imposing in our setup the hierarchy $\rho_1\ll\rho_T$), and the corresponding bound $m_\chi < 10$ GeV that follows from the requirement of the correct relic abundance. In this and the following section, we discuss the detection bounds.

The energy-momentum tensor for a Dirac field with mass $m$ is given by
\be
T^{\mu\nu}=\partial^\nu \bar\psi \frac{\partial \mathcal L}{\partial(\partial_\mu \bar\psi)}+\frac{\partial\mathcal L}{\partial(\partial_\mu\psi)}\partial^\nu\psi-\eta^{\mu\nu}\mathcal L\quad\textrm{with} \quad  \mathcal L=\frac{i}{2}\bar\psi \gamma^\mu\partial_\mu \psi-\frac{i}{2}\partial_\mu\bar\psi \gamma^\mu \psi-m\bar\psi\psi  \,.
\ee
Therefore, on-shell (after applying the Dirac equation on on-shell fermions) one gets
\be
T^{\mu\nu}=\frac{i}{2}\bar\psi\gamma^{(\mu} \partial^{\nu)}\psi-\frac{i}{2}\partial^{(\nu}\bar\psi\gamma^{\mu)}\psi=i\bar\psi\gamma^{(\mu} \partial^{\nu)}\psi,\quad \eta_{\mu\nu}T^{\mu\nu}=T=m\bar\psi\psi  \,,
\label{eq:emt}
\ee
where the brackets indicate symmetrization between the involved indices, and in the last equality we have integrated by parts and used that the coupling is with a symmetric and transverse tensor $h_{\mu\nu}(z_b,x)$.
In particular the energy momentum $T_\chi^{\mu\nu}$ and $T_Q^{\mu\nu}$ is given by Eq.~(\ref{eq:emt}) with the corresponding fermion fields $\chi$, with mass $m=m_\chi$, and $Q$, with $m=m_Q$.

\subsection*{Graviton mediation}
For KK graviton mediation the effective Lagrangian is given by Eq.~(\ref{eq:Leffgrav}), which becomes now
\be
\mathcal L_{\rm eff}=-\sum_{n=1}^\infty \left\{ a_n \left(\bar\chi\gamma^{\mu} \partial^{\nu}\chi \right) \left(\bar Q\gamma_{\mu} \partial_{\nu}Q\right)+b_n(\bar\chi \chi)(\bar QQ)
\right\} \,,\quad b_n=\frac{a_n}{3}m_\chi m_Q   \,.
\label{eq:LeffgravchiQ}
\ee

As we are considering only dimension 6 operators, we will disregard the first term in Eq.~(\ref{eq:LeffgravchiQ}), and we will concentrate on the second term which is a dimension 6 operator. As we will see, this term is subleading with respect to the similar effective Lagrangian induced by the exchange of the radion. For instance for $n=1$ its Wilson coefficient is given by
\be
b_1=0.1\, (k/M_{\rm Pl})^{-2}\,\frac{m_\chi m_Q}{\tilde\rho_T^4} \,.
\label{eq:b1}
\ee

\subsection*{Radion mediation}
For radion mediation the effective Lagrangian is given by Eq.~(\ref{eq:Leffrad}), which becomes now
\be
\mathcal L_{\rm eff}=b_r (\bar\chi \chi)(\bar Q Q) \, \quad \textrm{with}\quad  b_r =a_r m_\chi m_Q=\frac{m_\chi m_Q}{6 m_r^2\tilde\rho_T^2}  \,,
\label{eq:LeffradchiQ}
\ee
and has a similar structure to the second term in Eq.~(\ref{eq:LeffgravchiQ}). In fact the radion contribution will dominate the graviton contribution provided that the condition
\be
m_r \ll 1.3 \,(k/M_{\rm Pl})\tilde\rho_T=1.3 \, \rho_T
\ee
holds, which is always true. Therefore from here on we will only consider the effective Lagrangian from the radion exchange in Eq.~(\ref{eq:LeffradchiQ}).

\subsection{Bounds from nuclear recoil}

We will assume that $\tilde\rho_T$ is $\mathcal O$(TeV), and for concreteness we will fix it as $\tilde\rho_T\simeq 1$ TeV, while $m_\chi\lesssim \rho_1$ will be considered as a free parameter in the mass range $0.5 \textrm{ GeV}\lesssim m_\chi\lesssim 10$ GeV. Finally $m_r\ll \rho_1$ (for the consistency of the effective theory) is also considered as a free parameter.

The quark level should be matched into nucleon $N$ level operators. In particular the matrix elements of the light quarks ($Q=u,d,s$) can be computed in chiral perturbation theory as~\cite{Agrawal:2010fh}
\be
\langle N |m_Q\bar QQ |N\rangle=m_N f^{(N)}_{T_Q}
\ee
for $N=p,n$ given by the proton ($p$) or neutron ($n$). As for the heavy quarks $Q=c,b,t$ they connect to the gluons inside the nucleon through a loop diagram, giving
\be
\langle N|m_Q\bar Q Q|N\rangle=\frac{2}{27}m_N  \left(1-\sum_{Q=u,d,s} f^{(N)}_{T_Q}\right)\,,
\ee
where $f^{(p)}_{T_u}=0.018(5)$, $f^{(p)}_{T_d}=0.027(7)$, $f^{(p)}_{T_s}=0.037(7)$ and  $f^{(n)}_{T_u}=0.013(3)$, $f^{(n)}_{T_d}=0.040(10)$, $f^{(n)}_{T_s}=0.027(7)$ \cite{Ellis:2018dmb} determine the WIMP coupling $f_N$ to nucleons, given by
\be
\frac{f_N}{m_N}=\sum_{Q=u,d,s} \frac{b_r}{m_Q}f^{(N)}_{T_Q}+\frac{2}{27}\left(1-\sum_{Q=u,d,s} f^{(N)}_{T_Q}  \right)\sum_{Q=c,b,t}\frac{b_r}{m_Q}  \,,
\ee
which, using (\ref{eq:LeffradchiQ}), gives for protons and neutrons
\be
\frac{f_N}{m_N}=c_N \frac{m_\chi}{6 m_r^2 \rho_T^2},\quad c_p\simeq 0.29,\quad c_n\simeq 0.28  \,.
\ee
Then, the DM-nucleon spin-independent total cross-section is given by
\be
\sigma_N=\frac{\mu^2_{N\chi}}{\pi} f_N^2  \,, 
\label{eq:sigmaN}
\ee
where $\mu_{N\chi}$ is the reduced DM-nucleon mass $\mu_{N\chi}=m_N m_\chi/(m_N+m_\chi)$.  

\begin{figure}[t]
\centering
\includegraphics[width=8.5cm]{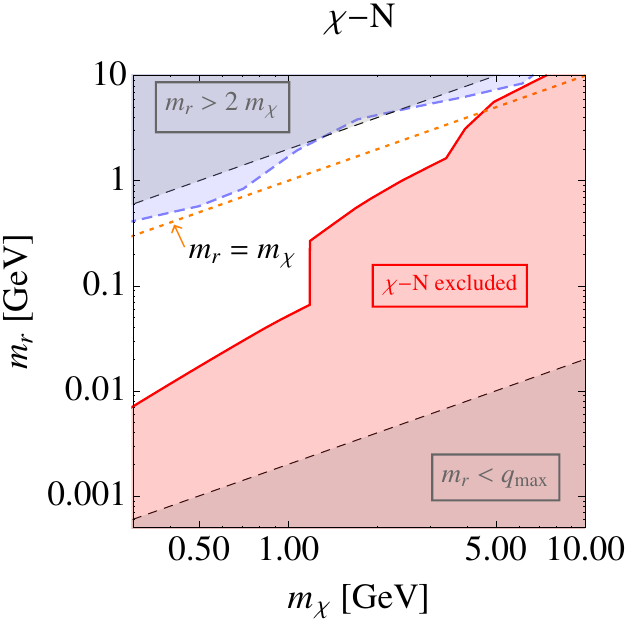} \hspace{0.5cm}
 \includegraphics[width=8.5cm]{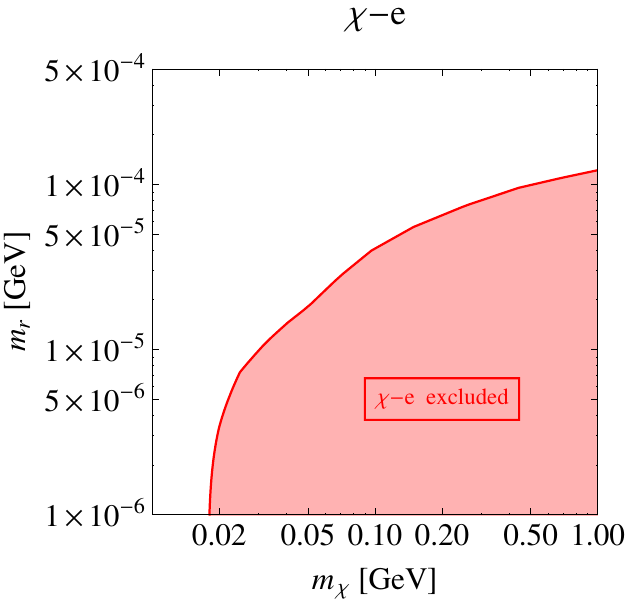} 
\caption{\it Left panel: Lower bound (red region) in the plane $(m_\chi,m_r)$ from the spin-independent DM-nucleon cross section, for heavy DM $m_\chi\gtrsim 0.3$ GeV, as a function of DM mass. The blue region stands for the neutrino coherent scattering limit. Data taken from Ref.~\cite{ParticleDataGroup:2022pth}.  We have considered $\tilde\rho_T = 1\,$TeV, and we have assumed $N =$ proton, although the neutron case would lead to an indistinguishable figure. In the region $m_r>2m_\chi$ the decay $r\to\chi\bar\chi$ can take place. Right panel: The same from the DM-electron scattering, for light DM $m_\chi\lesssim 1$ GeV.}
\label{fig:radion_bounds}
\end{figure}

Eq.~(\ref{eq:sigmaN}) can be compared with the experimental limits on direct detections. In the range $3\textrm{ GeV}\lesssim m_\chi\lesssim 10 \textrm{ GeV}$ the strongest bounds are given by XENON1T~\cite{XENON:2019zpr,XENON:2019gfn}, in the range $1\textrm{ GeV}\lesssim m_\chi\lesssim 3 \textrm{ GeV}$ by DarkSide50~\cite{DarkSide-50:2022qzh} and for $0.5\textrm{ GeV}\lesssim m_\chi\lesssim 1 \textrm{ GeV}$ by CREST~\cite{CRESST:2019jnq}. The dependence $\sigma_N \propto 1/m_r^4$ implies that for a given value of $m_\chi$ the experimental upper limits set a lower bound for the radion mass. This bound is displayed in the pink region in Fig.~\ref{fig:radion_bounds} (left panel). 

The blue region corresponds to the upper bound imposed by the neutrino floor to be detected by direct searches. The region inside the blue region is not excluded, but direct detection there seems most problematic as it is inside the neutrino floor. Values of the radion mass $m_r \lesssim m_\chi$ 
turn out to be compatible with the current experimental limits when considering $m_\chi \lesssim 2 \, \textrm{GeV}$.

Notice that the actual value of $\rho_1$ (or similarly the value of $\tilde\rho_1$) is not involved in Fig.~\ref{fig:radion_bounds}, only the value of $m_\chi$, which is naturally smaller than $\rho_1$. Moreover in this paper we are considering the region where $m_r\lesssim m_\chi$, so that in particular the decay of the radion into $\chi\bar\chi$ cannot take place. This region is favored by direct measurements, as we can see from the left panel of Fig.~\ref{fig:radion_bounds} where the region $m_r>2m_\chi$ (for which the decay $r\to\chi\bar\chi$ could take place) is inside the neutrino floor area and thus inaccesible to future experimental data. In this way considering the region $\rho_1\lesssim 3 $ GeV, in agreement with data from PTA experiments, an absolute upper bound can be set on the radion mass as $m_r\lesssim 3$ GeV.

Finally, on top of the lower bound from the experimental data on direct detection in Fig.~\ref{fig:radion_bounds}, there is another theoretical lower bound from the validity of the effective field theory we are using in Eq.~(\ref{eq:LeffradchiQ}). Direct detection is through the reaction $\chi(p)+\mathcal N(k)\to \chi(p')+\mathcal N(k')$, where $\mathcal N(k)$ is a nucleus, initially at rest, and $\chi(p)$ the non-relativistic dark matter, with respective momenta 
\be
k=(m_{\mathcal N},\vec 0),\quad p=\left(m_\chi+\frac{{\bf p}^2}{2m_\chi}, \vec p\right),\quad \textrm{with}\quad \vec p=m_\chi \vec v_\chi  \,,
\ee
where $\textbf p\equiv |\vec p|$ and $\textbf v_\chi\simeq 10^{-3}$.
Given the momentum transfer $q=p-p'=k'-k$, energy conservation implies that
\be
\frac{\textbf p\, \textbf q \cos\theta}{m_\chi}=\frac{\textbf q^2}{2\mu_{\chi\mathcal N}}  \,,
\label{eq:relacion}
\ee
where $\theta$ is the scattering angle in the laboratory frame.  From Eq.~(\ref{eq:relacion}) we get the maximum value for the momentum transfer $\textbf q$ as 
\be
\textbf q_{\rm max}=\frac{2\,\textbf p\,\mu_{\chi\mathcal N}  }{m_\chi}=2\mu_{\chi\mathcal N}\textbf v_\chi\simeq 2 m_\chi \textbf v_\chi  \,,
\ee
where in the last equality we have used $m_\mathcal N \gg m_\chi$. Therefore the validity of the effective theory is guaranteed provided that $m_r>\textbf q_{\rm max}\simeq 2\times 10^{-3}m_\chi$. The region where this condition is not satisfied is shown in the left panel of Fig.~\ref{fig:radion_bounds}, which shows that the region where the effective theory is not valid was already excluded by direct measurements, so it has no impact in the model bounds.

\subsection{Sub-GeV Dark Matter}
Dark matter detection via nuclear recoil becomes much less effective for sub-GeV DM masses due to the much smaller recoil energies. This will happen in particular for the region where the first order confinement/deconfinement phase transition in the $\mathcal B_1$ brane generates the stochastic gravitational waves background that could be detected by PTA collaborations, Ref.~\cite{Megias:2023kiy}, \textit{i.e.}~the region  $\rho_1\in[10 \textrm{ MeV},10 \textrm{ GeV}]$, which implies from the right panel of Fig.~\ref{fig:Tr} that $m_\chi\lesssim 2$ GeV. 

One possible avenue that has been undertaken is to replace the nucleon recoil by electron recoil in an atomic target. The electron is in a bound state with a typical wave function size $R_{\rm Bohr}=1/\alpha m_e$ and a typical momentum $\sim 1/R_{\rm Bohr}=\alpha m_e$. A rough approximation is considering scattering off free electrons $\chi(p)+e(k)\to \chi(p')+e(k')$, and given that the electron velocity $v_e\sim \alpha$ is much larger than the DM velocity $v_\chi$, kinematics leads to the momentum transfer $q$, with $q^0\ll\mathbf{q}\simeq \mu_{\chi e} v_e\simeq m_e\alpha$, for $m_\chi\gg m_e$. For the previous scattering process the spin-averaged matrix element $|\mathcal M|^2$ is given by
\be
|\mathcal M|^2=\frac{m_\chi^2 m_e^2}{9\tilde\rho_T^4}\frac{(m_\chi^2-q^2/4)(m_e^2-q^2/4)}{(q^2-m_r^2)^2}\simeq \frac{m_\chi^2 m_e^2}{9\tilde\rho_T^4}\frac{(m_\chi^2+\textbf{q}^2/4)(m_e^2+\textbf{q}^2/4)}{(\textbf{q}^2+m_r^2)^2}\simeq \frac{m_\chi^2 m_e^2}{9\tilde\rho_T^4}\frac{m_\chi^2m_e^2}{(\textbf{q}^2+m_r^2)^2}\,,
\ee
where, in the last equality we are using that $\mathbf q^2\simeq \alpha^2 m_e^2\ll m_e^2\ll m_\chi^2$. 
Following the model independent approach of Ref.~\cite{Essig:2015cda} we parametrize the elastic scattering cross section $\bar\sigma_e$ as
\be
\frac{\mu^2_{\chi e}}{16\pi m_\chi^2 m_e^2}\left| \mathcal M(\textbf{q}) \right|^2\equiv  \bar\sigma_e F_{\rm DM}^2(\mathbf q),\quad F_{\rm DM}(\mathbf q)=\frac{\alpha^2 m_e^2+m_r^2}{\mathbf q^2+m_r^2}  \,.
\ee

A contour plot of the upper bounds on $\bar\sigma_e$ in the plane $(m_\chi,m_r)$, from experimental measurements, is shown in the right panel of Fig.~\ref{fig:radion_bounds}. The value of $\bar\sigma_e$ is constrained by experimental results which are putting upper bounds on it depending on the DM mass. In particular for $50 \textrm{ MeV}\lesssim m_\chi\lesssim 1$ GeV it is mainly constrained by XENON1T data~\cite{XENON:2021qze}, for 
$15 \textrm{ MeV}\lesssim m_\chi\lesssim 50 \textrm{ MeV}$ by DarkSide data~\cite{DarkSide:2022knj}, and for $1 \textrm{ MeV}\lesssim m_\chi\lesssim 15 \textrm{ MeV}$ by SENSEI at SNOLAB data~\cite{SENSEI:2023zdf}.
For $m_r > \alpha m_e$ the average cross section is approximated by
\be
\bar\sigma_e\simeq \frac{\mu^2_{e\chi}}{144\pi\tilde\rho_T^4}\frac{m_\chi^2m_e^2}{m_r^4}  \,,
\ee
and the considered experimental data are those for which the form factor with $F_{\rm DM}(\mathbf q)\simeq 1$ does apply. This is the region shown in the right panel of Fig.~\ref{fig:radion_bounds}.
On the other hand,  for $m_r<\alpha m_e$ the average cross section is approximated by
\be
\bar\sigma_e\simeq \frac{\mu^2_{e\chi}}{144\pi\tilde\rho_T^4}\frac{m_\chi^2}{\alpha^4 m_e^2}   \,,
\ee
and the corresponding experimental data are those for which the form factor $F_{\rm DM}(\mathbf q)\simeq \alpha m_e/\mathbf q^2$ applies. In this region, not shown in Fig.~\ref{fig:radion_bounds}, the experimental result do not impose any constraints in the parameter space of our model.

\section{Accelerator searches}
\label{sec:accelerator}
The most promising dark matter searches at the LHC in our model are in events with missing energy and a $Z$ boson~\cite{Carpenter:2012rg}. This analysis is based on an effective Lagrangian description between quarks and DM, which in our model is given by
\be
\mathcal L=\frac{m_q}{\Lambda^3}(\bar q q)(\bar\chi\chi),\quad \textrm{where}\quad \Lambda=\left( \frac{6m_r^2\trho_T^2}{m_\chi}\right)^{1/3}  \,.
\ee
Using now data from ATLAS on $ZZ$ production~\cite{ATLAS:2012bra} a lower bound on $\Lambda$ is found as a function of $m_\chi$. In the region $1\,\textrm{GeV}\lesssim m_\chi\lesssim 10$ GeV the 90\% CL bound $\Lambda\gtrsim 0.1$ GeV is found~\cite{Carpenter:2012rg}. For the model we are considering in this paper it translates into the bound
\be
m_r\gtrsim 10^{-5}\, \textrm{GeV} \left(\frac{m_\chi}{1\,\textrm{GeV}} \right)^{1/2}  \,,
\ee
which is easily satisfied and consistent with all other constraints.

In the mass range $m_e\lesssim m_r\lesssim m_p$ the fixed-target experiments can provide the advantage of high-energy particle beams and relatively large intensities: in particular the NA64 experiment at CERN SPS~\cite{NA64:2023wbi} and the future LDMX experiment at SLAC~\cite{Berlin:2018bsc}. Here we can apply the search for a new generic boson, the radion in our case, particle produced in the 100 GeV electron scattering off nuclei $(A,Z)$, $e^-Z\to e^- Zr$, followed by its invisible decay in the NA64 experiment at CERN. Defining the coupling of the radion to electron
\be
\mathcal L=-g_{ree}r\,\bar e\, e ~~ \quad {\rm with} \quad ~~ g_{ree}=\frac{m_e\trho_1}{\sqrt{6}\trho_T^2}  \,,
\ee
and NA64 data~\cite{NA64:2021xzo,NA64:2023wbi}, one can put upper bounds on the value of the coupling $g_{ree}$: $g_{ree}\lesssim 5\cdot 10^{-6}$ for $m_r=10^{-3}$ GeV, and $g_{ree}\lesssim 3\cdot 10^{-3}$ for $m_r=1$ GeV~\cite{NA64:2021xzo}. In our case
\be
g_{ree}= 2\times 10^{-10}\left( \frac{\trho_1}{1\,\textrm{GeV}} \right)<2\times 10^{-9},\quad \textrm{for}\quad \trho_1<10\,\textrm{GeV}  \,,
\ee
well below the experimental bounds. 

Notice that the previous bounds are based on the assumption that the radion field decays invisibly, \textit{i.e.}~in the channel $r\to \chi\bar\chi$, which is kinematically allowed if $m_r>2m_\chi$, with a width
\be
\Gamma_{r\to\chi\bar\chi}=\frac{m_r m_\chi^2}{48\pi\trho_1^2}\left(1-\frac{4m_\chi^2}{m_r^2}  \right)^{3/2}\,.
\ee
Otherwise, if $m_r<2m_\chi$, the invisible channel is forbidden and the radion decays into SM particles. In particular it will decay into the SM fermions $f$, such that $m_r>2 m_f$, with a width
\be
\Gamma_{r\to f\bar f}=  N_c\frac{m_r \trho_1^2 m_f^2}{48\pi\trho_T^4}\left( 1-\frac{4m_f^2}{m_r^2}\right)^{3/2}   \,, 
\label{eq:rtoff}
\ee
while lighter fermion contributions are highly suppressed. For values of $m_r\gtrsim \Lambda_{\rm QCD}$, \textit{i.e.}~in the perturbative regime of the strong coupling, the radion can also decay into a couple of gluons $r\to gg$ with a width
\be
\Gamma_{r\to gg}=\frac{\alpha_3^2 b_{\rm QCD}^2}{192 \pi^3 }\frac{m_r^3 \trho_1^2}{\trho_T^4}  \,,
\label{eq:rtogg}
\ee
while the decay channel into photons $r\to\gamma\gamma$ 
\be
\Gamma_{r\to \gamma\gamma}=\frac{\alpha_{\rm QED}^2 b^2_{\rm QED}}{1536\pi^3}\, \frac{m_r^3\tilde\rho_1^2}{\tilde\rho_T^4} \,,
\label{eq:rtogammagamma}
\ee
has a much smaller width, and will be neglected.

\begin{figure}[t]
\centering
 \includegraphics[width=10cm]{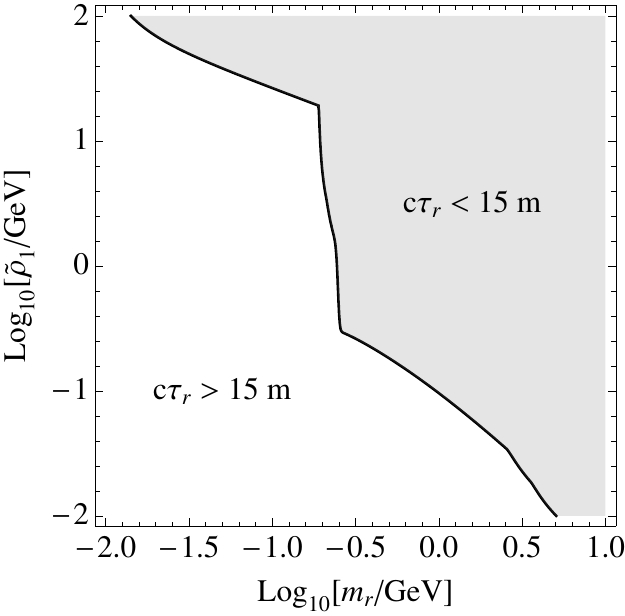} 
\caption{\it Contour line of $c\tau_r=15\,$m in the plane $(m_r,\trho_1)$. Below the solid line $c\tau_r>15\,$m.}
\label{fig:radion_decay}
\end{figure}

In the region $m_r<2m_\chi$ the radion decays into SM fields with a width $\Gamma_r\simeq \sum_f \Gamma_{r\to f\bar f}+\Gamma_{r\to gg}$ and a lifetime $\tau_r=1/\Gamma_r$. If the radion decays inside the detector, the previous bounds are not valid as they are obtained under the assumption of an invisible decay of the radion and the searches are based on detection of missing energy. However if the radion decays outside the detector, there is no difference from the detection point of view, with the case where it decays invisibly, as there will be missing energy inside the detector and the bounds do apply. As the decay volume in the NA64 detector is a cylinder of 30~cm diameter and 15~m length, we can conservatively impose the condition that $c\tau_r>15$ m, for the radion to decay outside the detector. The region of the parameters in the plane $(m_r,\trho_1)$ is exhibited in Fig.~\ref{fig:radion_decay}.
Below the solid line in Fig.~\ref{fig:radion_decay} the radion decays outside the detector. 

Finally notice that, using the rough bounds $m_f<m_r/2<\trho_1/2<50$ GeV we can set the absolute bound $\Gamma_r/m_r < 10^{-7}$, although in most of the considered parameter space $\Gamma_r/m_r$ can be orders of magnitude smaller, which justifies neglecting  the width $\Gamma_r$ in all previous calculations where the radion is propagating.

\section{Indirect constraints}
\label{sec:indirect}
DM annihilation can leave observable traces today and over the early history of the universe. As $\chi\bar\chi$ annihilates into leptons, quarks, neutrinos and massless gauge bosons, we can now compare the model predictions with a number of experimental data measured today and in the early universe. As a general trend, indirect detection puts strong lower bounds on the DM mass when the detected annihilation channels coincide with the annihilation channels giving rise to the DM cosmic abundance. However, as we will see, in our model annihilation channels involved in indirect detection are very inefficient and do not involve the annihilation channel, \textit{i.e.}~$\chi\bar\chi\to rr$, giving rise to the relic abundance, allowing to avoid bounds from indirect detection.

In the following, we first study the cosmological evolution of the $(\chi,r)$ system interacting with the thermal bath of the SM particles by solving the coupled Boltzmann equations. Among other things, this is a useful verification of the relic abundance calculation of the fermion $\chi$, presented in Sec.~\ref{sec:abundance}.

\subsection{Radion cosmology and bounds from Big Bang Nucleosynthesis}
In our model we are considering a light radion, which can decay into SM channels whenever its decay rate $\Gamma_r$ is larger than the expansion rate of the universe, i.e.~$\Gamma_r>H$.
The SM fields are in equilibrium with the thermal plasma through gauge and Yukawa interactions. The last fermion to come to thermal equilibrium is $e_R$, which comes into equilibrium with $e_L$ at the temperature $T \sim 80$ TeV~\cite{Joyce:1997uy} due to the tiny Yukawa coupling, via the reactions $h \leftrightarrow e_Le_R$.

The Boltzmann equation for the radion number density $n_r$ and the DM number density $n_\chi$ can be written as
\begin{align}
\frac{dn_r}{dt}+3Hn_r &= - \langle \Gamma_r \rangle\left[n_r-n_r^{\rm eq}\right] -\langle \sigma_r v\rangle[n_r^2-(n_r^{\rm eq})^2]+\langle \sigma_\chi v\rangle \left[ n_\chi^2-\left(\frac{n_\chi^{\rm eq}}{n_r^{\rm eq}} \right)^2n_r^2\right] \,, \nonumber\\
\frac{dn_\chi}{dt}+3Hn_\chi &=-\langle \sigma_\chi v\rangle \left[ n_\chi^2-\left(\frac{n_\chi^{\rm eq}}{n_r^{\rm eq}} \right)^2n_r^2\right]-\langle \sigma_{0} v\rangle[n_\chi^2-(n_\chi^{\rm eq})^2] \,,
\label{eq:B1}
\end{align}
where $\Gamma_{r}\equiv \Gamma(r\to$ SM+SM) $\sim c_r^2$ is the decay width defined in Eqs.~(\ref{eq:rtoff}), (\ref{eq:rtogg}) and (\ref{eq:rtogammagamma})~\footnote{$\langle \Gamma_r\rangle$ stands for the thermally averaged (time dilation included) decay width of the radion, which is computed as
\begin{equation}
\langle \Gamma_r\rangle = \frac{\sum_{n=1}^\infty \frac{1}{n}  K_1(nx)}{\sum_{n=1}^\infty \frac{1}{n} K_2(nx)} \Gamma_r  \,,
\end{equation}
where $\Gamma_r$ is the decay width of the radion at rest.},
while the different cross-sections are defined as: $\sigma_r\equiv\sigma(rr\to$ SM+SM) $\sim c_r^4$, $\sigma_\chi\equiv \sigma(\chi\bar\chi\to rr)\sim c_r^0$ and $\sigma_{0}\equiv \sigma(\chi\bar\chi\to$ SM+SM) $\sim c_r^2$, where we have indicated the respective orders in the small parameter $c_r(z_T)$. 
Eq.~(\ref{eq:B1}) can be written changing variable to $Y_r=n_r/s$ and $x=m_r/T$ as
\begin{align}
\frac{dY_r}{dx}&=-\gamma x\left[ Y_r-Y_r^{\rm eq}\right]-\frac{\lambda_r}{x^2} [Y_r^2-(Y_r^{\rm eq})^2] +\frac{\lambda_\chi}{x^2}\left[ Y_\chi^2-\left(\frac{Y_\chi^{\rm eq}}{Y_r^{\rm eq}} \right)^2 Y_r^2 \right] \,, \nonumber\\
\frac{dY_\chi}{dx}&=-\frac{\lambda_\chi}{x^2}\left[ Y_\chi^2-\left(\frac{Y_\chi^{\rm eq}}{Y_r^{\rm eq}} \right)^2  Y_r^2\right]-\frac{\lambda_{0}}{x^2}[Y_\chi^2-(Y_\chi^{\rm eq})^2  ] \,, \nonumber\\
 \gamma
& \simeq \gamma_0 \frac{K_1(x)}{K_2(x)},\quad \gamma_0\equiv\frac{\Gamma_r}{H(m_r)}, \quad \lambda_i=\frac{s(m_r)\langle \sigma_i v\rangle}{H(m_r)}, \quad (i=r,\chi,0) \,,
\label{eq:B2}
\end{align}
where $H(m_r)$ is the Hubble constant at the temperature $T=m_r$, $Y_{r,\chi}^{\rm eq}$ are given by Eq.~(\ref{eq:nchi}),
and the cross-sections $\sigma_{i}$ $(i=r,\chi)$ are $p$-wave suppressed so that $\sigma_i=\sigma_{i}^0 v$, and $\langle \sigma_{i} v\rangle=2\sigma_{i}^0+\mathcal O(x^2)$ for $x\ll 1$. Taking into account the dependence of the different couplings on the small parameter $c_r(z_T)$, we find the following hierarchy among them: $\lambda_r\ll \lambda_{0},\, \gamma_0\ll \lambda_\chi$.
For instance we can consider the different cross-section and decay rates into SM fermions $f$. In this case we find that, parametrically in terms of the couplings
\be
\frac{\lambda_0}{\lambda_\chi}\simeq g_{rff},\quad \frac{\gamma_0}{\lambda_\chi}\simeq g_{rff},\quad \frac{\lambda_r}{\lambda_\chi}\simeq g_{rff}^2,\quad
g_{rff}=\frac{m_f\rho_1}{\sqrt{6} \rho_T^2}
\ee
where e.g.~for $f=e$ the coupling is typically $g_{ree}\simeq 10^{-10}$. Similar suppression is obtained for processes into photons and/or gluons.

\begin{figure}[t]
  \centering
\includegraphics[width=7cm]{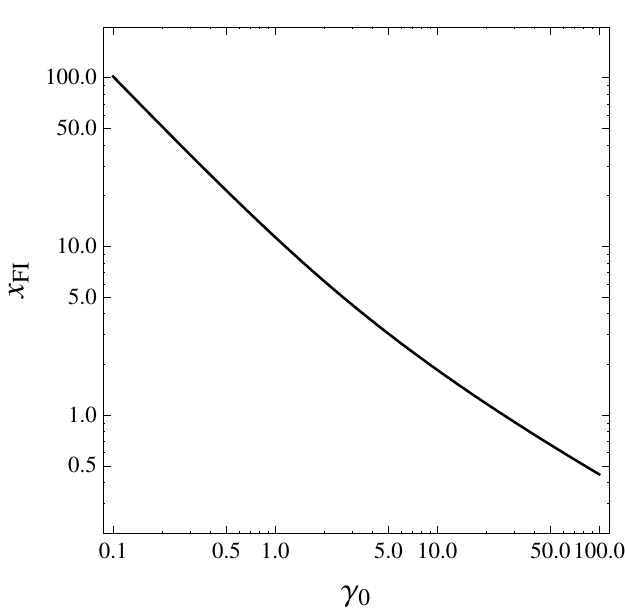}  
\caption{\it The freeze-in temperature $x_{\rm FI}$ as a function of $\gamma_0$.}
\label{fig:xFI}
\end{figure} 
Some analytical properties of Eq.~(\ref{eq:B2}) are in order before exploring it numerically. For very high temperatures ($x\ll 1$), for which $\lambda_{i}/x^2\gg 1$, the equilibrium solution $Y_r=Y_r^{\rm eq}$ and $Y_\chi=Y_\chi^{\rm eq}$ follows. Concerning the radion abundance, for a value of the temperature $T_{r,\FO} \simeq m_\chi$, corresponding to $x_{r,\FO} = m_r/T_{r,\FO}  < 1$, such that $\lambda_r/x_{r,\FO}^2 \lesssim \mathcal O(1)$ (i.e.~$x_{r,\FO} \gtrsim \lambda_r^{1/2}$), the radion goes out of equilibrium while $\chi$ stays in equilibrium as $\lambda_\chi/x_{r,\FO}^2 \gg 1$. One can easily check using numerical analysis  that the solution we find is an attractor, which does not depend on the initial conditions as the equilibrium solution for $Y_r=Y_r^{\rm eq}$ is quickly recovered~\footnote{This behavior, with re-equilibration at low temperature, was already observed  for the case of ALPs $a$ from the three-point function $a\leftrightarrow \gamma\gamma$~\cite{Millea:2015qra}.}, while a constant out of equilibrium solution $Y_\chi=$ constant is found for the $\chi$ distribution. In fact one can define a value of $x=x_{\rm FI}$ at which $Y_r\simeq Y_r^{\rm eq}$ by the condition $\gamma \, x_{\rm FI}  \simeq \mathcal O(10)$ (where we are conservatively considering one order of magnitude) which fixes a freeze-in temperature $T_{\rm FI}$ for the radion to re-enter the equilibrium distribution. The thermal history for the radion is then the following: \textit{i)} it makes part of the SM plasma at very high temperatures until it freezes-out from the SM at $T_{r,\FO}$, and \textit{ii)} finally it re-enters thermal equilibrium with the SM plasma at the temperature $T_{\rm FI}$.

A plot of $x_{\rm FI}$ as a function of $\gamma_0$ is provided in Fig.~\ref{fig:xFI} from the condition $\gamma\, x_{\rm FI} = 10$. 
We now have two extreme possibilities: 
\begin{itemize}
\item
If $\gamma_0\gg1$, then  $x_{\rm FI}\simeq \sqrt{20/\gamma_0}\ll1$.
\item
If $\gamma_0\ll 1$, then $x_{\rm FI}=10/\gamma_0\gg 1$.
\end{itemize}

We can  then \textit{a posteriori} validate our previous result on the value of $\Omega_\chi h^2$ provided that $T_{\rm FI}> T_{\rm FO}$: $Y_\chi$ will go to a constant value at the freeze-out temperature $T_{\rm FO}$, obtained in previous sections under the assumption that $Y_r=Y_r^{\rm eq}$. 
In fact this  condition can be implemented by imposing that the parameter $\sigma\gtrsim 1$, where $\sigma$ is defined as
\be
\sigma\equiv\frac{T_{\rm FI}}{T_{\rm FO}}\approx \frac{24 m_r}{x_{\rm FI}(\gamma_0)m_\chi} \,.
\label{eq:sigma}
\ee

\begin{figure}[t]
\centering
\includegraphics[width=8cm]{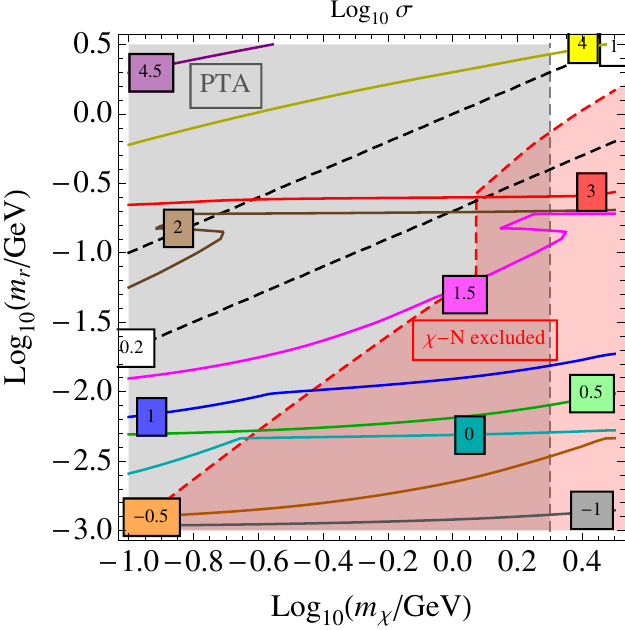} \quad \includegraphics[width=8cm]{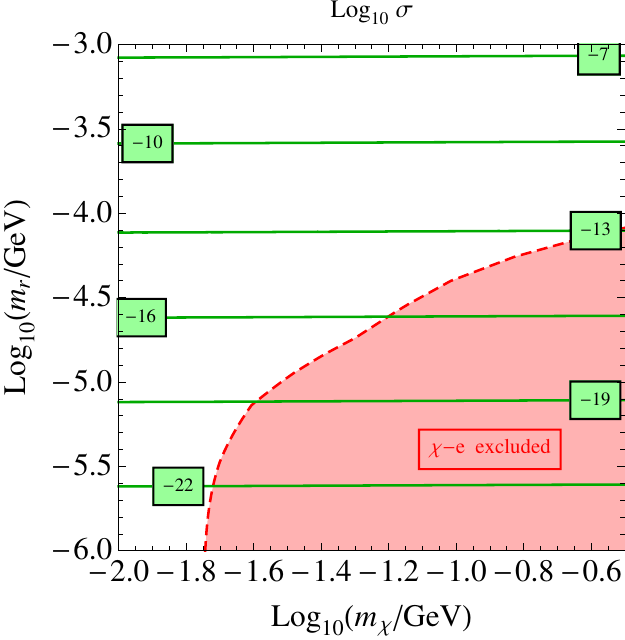}
\caption{\it Left panel: Contour lines of $\sigma$ for $m_r>1$ MeV. We also display the area of $m_\chi\lesssim 2 \, \textrm{GeV}$ where PTA results can be reproduced (shadowed gray area), and the forbidden region for the spin-independent DM-nucleon cross section (shadowed red area). The dashed straight lines correspond to fixed values of $m_r/m_\chi$. Right panel: Contour plots of $\sigma$ for $m_r<1$ MeV.  The forbidden region for the DM-electron collision is also displayed (shadowed red area).} 
\label{fig:sigma}
\end{figure}

A contour line plot of $\sigma$ is shown in Fig.~\ref{fig:sigma} for $m_r>1$ MeV (left panel) and for $m_r<1$ MeV (right panel). 
We can see that for $m_r>1$ MeV (left panel) the condition $\sigma\gtrsim 1$ is fulfilled, while for $m_r<1$ MeV (right panel) it turns out that $\sigma\ll 1$.  Furthermore,  it is interesting to notice that,  as we have checked in the numerical analysis, for $m_r > 1$ MeV  (that is for $T_{\FI}>T_{\FO}$) the value of $x$ at which $Y_\chi$ stabilizes is  insensitive to the value of $\sigma$ and the whole region, where $m_r>1$ MeV, can provide a good value of the relic density $\Omega_\chi h^2$.  This is illustrated in Fig.7. This is because 
in that range of $m_r$,  the dark matter $\chi$ is in thermal equilibrium through its interaction with the radions,  due to a large cross section $\sigma_\chi$.  Finally, at the temperature $T_{\rm FO}$, $\chi$ will freeze-out  and will go out of equilibrium with radions, stabilizing the value of $Y_\chi$.  Concerning the $m_r<1$ MeV region, as we will see,   is almost excluded by BBN conditions.

Notice that, for $m_r>1$ MeV, from the left panel of Fig.~\ref{fig:sigma} it follows that $T_{\rm FI}\gtrsim T_{d}$, being $T_d$ the temperature at which $\chi$ and the SM go out of equilibrium, i.e.~for which $\lambda_0/x^2(T_d)\ll 1$. Still as $\lambda_\chi\gg \lambda_0$ the parameter $\lambda_\chi/x^2(T_d)$ is not negligible, the density $Y_\chi$ does not stabilize yet at $T=T_d$ and decreases roughly as its equilibrium distribution, until the freeze-out temperature $T_{\rm FO}<T_d$ for which $\lambda_\chi/x^2(T_{\rm FO})\ll 1$. Then at the temperature $T_{\rm FO}$ the distribution $Y_\chi$, provided by Eq.~(\ref{eq:B2}), goes to a constant value $\sim 2 x^2(T_{\rm FO})/\lambda_\chi$.

To solve numerically the system (\ref{eq:B2}) we can make some approximations. Given that $\lambda_r\ll \lambda_\chi$ and $\lambda_0\ll \lambda_\chi$ one can safely neglect the corresponding terms in (\ref{eq:B2}). Moreover even if the cross-section $\sigma_\chi$ is $p$-wave suppressed, the inequality $\lambda_0\ll \lambda_\chi$ holds as, for the cross-section $\sigma\simeq a+b\, v$ we can write  $\langle \sigma v\rangle\simeq a+6b/x$ and, given that the freeze-out temperature is $T_{\rm FO}\simeq m_\chi/24$ and $x(T_{\rm FO})\simeq 24\, m_r/m_\chi$, the $p$-wave suppression $\lambda_\chi$ is much milder than the $\lambda_0$ suppression from the factor $c_r(z_T)$. We then can write $\lambda_\chi\equiv \lambda_\chi^0/x$ and approximate the Boltzmann equations as
\begin{align}
\frac{dY_r}{dx}&=-\gamma x\left[ Y_r-Y_r^{\rm eq}\right] +\frac{\lambda_\chi^0}{x^3}\left[ Y_\chi^2-\left(\frac{Y_\chi^{\rm eq}}{Y_r^{\rm eq}} \right)^2 Y_r^2 \right] \,, \nonumber\\
\frac{dY_\chi}{dx}&=-\frac{\lambda_\chi^0}{x^3}\left[ Y_\chi^2-\left(\frac{Y_\chi^{\rm eq}}{Y_r^{\rm eq}} \right)^2  Y_r^2\right] \,.
\label{eq:B2simp}
\end{align}
The relation between the constant value of $Y_\chi$ when it freezes-out, and the relic density, is given by
\be
\Omega_\chi=\frac{\rho_{\chi,0}}{\rho_{\rm crit,0}}=\frac{m_\chi Y_\chi s_0}{3M_P^2 H_0^2} \,,
\ee
where $s_0=(2\pi^2/45)g_{S,0}T_0^3$ is the entropy density today, with today's entropy number of degrees of freedom $g_{S,0}=3.94$, $T_0\simeq 2.4 \times 10^{-13}$ GeV the universe temperature today, and $H_0\simeq 2.14\times 10^{-42}\, h$ GeV the Hubble constant today. Putting numbers we get
\be
\Omega_\chi h^2\simeq \frac{3.5\times 10^{11}}{\lambda_\chi^0}\left(\frac{m_r}{\rm GeV}\right) \left(\frac{m_r}{m_\chi} \right) \,.
\ee

\begin{figure}[t]
\centering
\includegraphics[width=8cm]{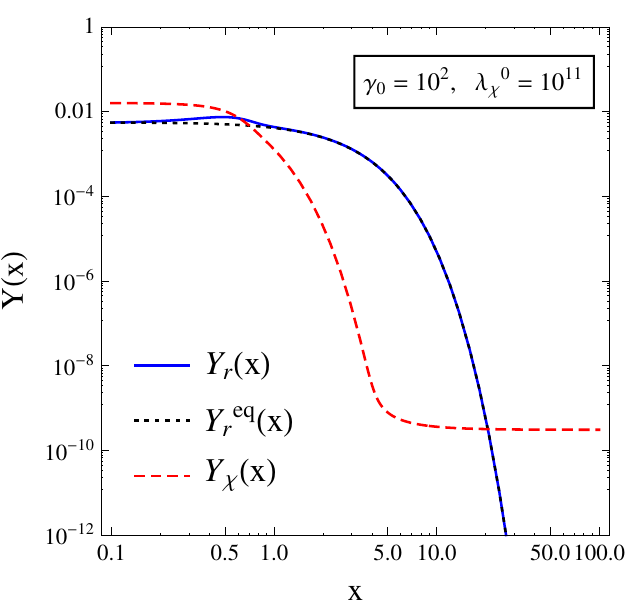} \quad \includegraphics[width=8cm]{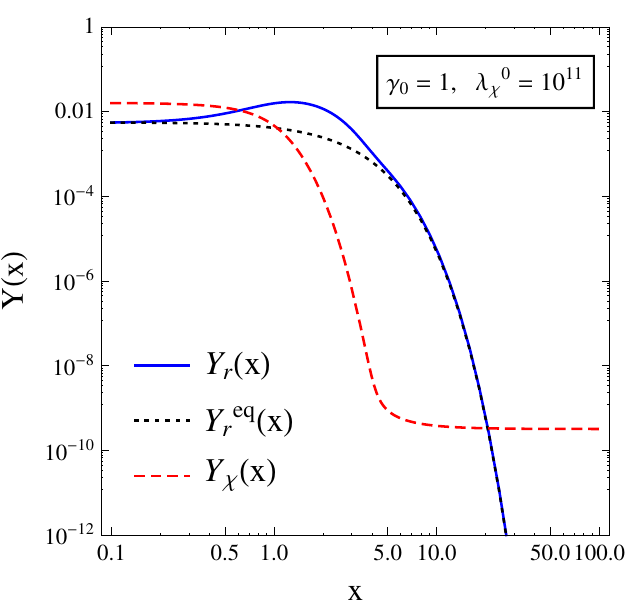}  \\
\caption{\it Left panel: Plots of $Y_r(x)$ and $Y_\chi(x)$ from Eq.~(\ref{eq:B2simp}) for $\gamma_0=10^2$ and $\lambda_\chi^0=10^{11}$. Right panel: The same as in the left panel for $\gamma_0 = 1$. We have considered $m_\chi = 1\,$GeV and $m_r = 0.2\,$GeV in both panels.}
\label{fig:numerical}
\end{figure}

The numerical solution of Eqs.~(\ref{eq:B2simp}) is provided in Fig.~\ref{fig:numerical} for $m_r=0.2$ GeV, $m_\chi=1$ GeV, $\gamma_0=10^2$, $\lambda_\chi^0=10^{11}$ (left panel), and  $\gamma_0=1$, $\lambda_\chi^0=10^{11}$
(right panel). As anticipated, we can see from the results of both panels of Fig.~\ref{fig:numerical} that the stabilizing value of the distribution $Y_\chi$ is not very sensitive to the actual value of $\gamma_0$, i.e.~on the value of $\sigma$. More details of this solution will be given elsewhere.

\subsubsection*{The region $m_r>2 m_e$}

\begin{figure}[t]
\centering
 \includegraphics[width=8cm]{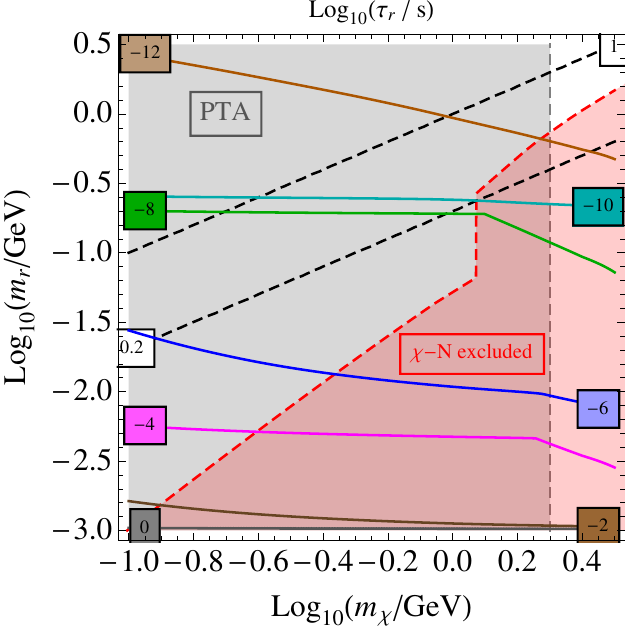}  \quad \includegraphics[width=8cm]{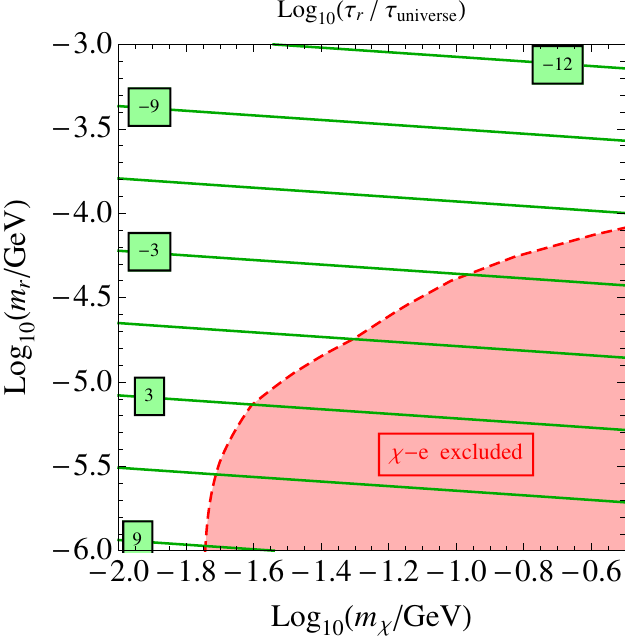}  
\caption{\it Left panel: Contour lines of $\tau_r/$sec for $m_r>1$ MeV. The dashed straight lines correspond to fixed values of $m_r/m_\chi$. Right panel: Contour lines of $\tau_r/\tau_{\rm universe}$ for  $m_r<1$ MeV.}
\label{fig:radion_lifetime}
\end{figure}

In the region $m_r>2 m_e\simeq 1$ MeV, the radion can decay into photon, gluon and fermion channels, the former being subleading in the total width $\Gamma_r$.  As we already have discussed,  the condition $m_r> 1$ MeV is sufficient for getting the correct relic abundance of $\chi$.
Moreover, BBN is not perturbed provided that $\tau_r\lesssim 10$ sec~\cite{Abu-Ajamieh:2017khi}. In particular, in the region where the radion width is dominated by the channel $r\to ee$, the radion lifetime is given by
\be
\tau_r\simeq 0.4 \,\textrm{sec} \left( \frac{\tilde \rho_T}{\textrm{TeV}} \right)^4 \left(\frac{\textrm{GeV}}{\tilde\rho_1}  \right)^2 \left(\frac{\textrm{MeV}}{m_r}  \right)   \,,
\ee
which easily satisfies the BBN bound. For heavier radion masses, where heavier fermions contribute to the decay $r\to f\bar f$, as the width is proportional to $m_r m_f^2$ the radion lifetime is shorter and the BBN bound is more easily satisfied, as it is shown in Fig.~\ref{fig:radion_lifetime} (left panel), from where it follows that the BBN constraint holds in the region where $m_r\gtrsim 1$ MeV. 

\subsubsection*{The region $m_r<2 m_e$}

For $m_r<2m_e$ the process $r\to\gamma \gamma$ is given in (\ref{eq:rtogammagamma}) for which $b_{\rm QED}\simeq \frac{7}{90} \frac{m_r^2}{m_e^2}$, which is zero to leading order for $m_r^2/m_e^2 \ll 1$, and we have only considered the leading effect from non-decoupling of  the electron. 

Contour lines of the parameter $\sigma$ are plotted in the right panel of Fig.~\ref{fig:sigma}, where it is shown that $\sigma\ll 1$. In that case the second term of the right-hand side of Eq.~(\ref{eq:B2simp}) gets larger than the equilibrium values, and the solution provided in previous sections should be revised. We plan to further study this situation in a different work. In this region we find that $x_{\rm FI}\gg 1$, or $T_{\rm FI}\ll m_r$,  so that processes as SM+SM $\to r$ are kinematically forbidden, preventing the radion from reentering thermal equilibrium.

In the region $m_r<1$ MeV the total decay width of the radion is dominated by the channel $r\to\gamma\gamma$ so that the BBN condition $\tau_r<10$ sec is never fulfilled, and thus it is excluded by the BBN constraint. Still the radion can be stable if its lifetime is larger than the universe lifetime, so it behaves as a stable particle, as it is shown in the right panel of Fig.~\ref{fig:radion_lifetime} which shows that in our model $\tau_r>\tau_{\rm universe}$ for $m_r\lesssim 10$ keV. In that case the radion could be an extra candidate to dark matter (see e.g.~Ref.~\cite{Abu-Ajamieh:2018brk} for earlier work). This possibility will be further pursued elsewhere. Nevertheless, we complete this section with a discussion of cosmological bounds on such light long-lived radions.

Radions keep thermal equilibrium until the freeze-out temperature $T_{r,\FO}$, at which they go out of equilibrium. The latter is expected to be $T_{r,\FO} \gg m_r$, as the coupling $\lambda_r$ is suppressed as $c_r^4$. Radions then will contribute to the effective number of neutrinos $\Delta N_{\rm eff}$, with an amount which depends on the actual value of the decoupling temperature $T_{r,\FO}$.
\newline\indent
If $T_{r,\FO}$ is such that $m_e < T_{r,\FO} < m_\mu$ (like the neutrinos), they
will contribute to the effective number of degrees of freedom as $\Delta N_{\rm eff}=4/7=0.57$, which is $\sim 3.4 \sigma$ away from the Planck value $N_{\rm eff}=2.99\pm 0.17$~\cite{Planck:2018vyg}.~\footnote{Of course, if $T_{0} < T_{r,\FO} <m_e$, this scenario would change the Standard Cosmology as the standard relation between the photon and neutrino temperature  $T_\gamma=\left( \frac{11}{4} \right)^{1/3}T_\nu$ would change to $T_\gamma=\left( \frac{13}{4} \right)^{1/3}T_\nu$ because the radion is in thermal equilibrium at $T=m_e$ and the radion heats photons by the reaction $rr\to\gamma\gamma$ at decoupling, which gives the relation $T_\gamma=\left(\frac{3}{2} \right)^{1/3}T_r$. Using Eq.~(\ref{eq:DeltaNeff}) leads to $\Delta N_{\rm eff}=\frac{4}{7} (13/6)^{4/3}\simeq 1.6$, a value flagrantly excluded by Planck data.}
This number gets smaller if radions go out of equilibrium before neutrinos decouple at $T_\nu\sim 1$ MeV, and in particular at temperatures higher than the mass of some SM particles. In this case,  when the temperature goes below the mass of the SM particle, the latter becomes non-relativistic and annihilates, so that neutrinos, photons, and the rest of relativistic SM particles, but not radions, would have been heated, and the contribution would be $\Delta N_{\rm eff}\ll 4/7$. The simplest example was given in Ref.~\cite{Weinberg:2013kea} for $T_{r,\FO} > m_\mu$, when neutrinos are still  in thermal equilibrium, but the decoupling temperature of the radion is below the mass of the other SM particles and/or scales. Then for $T\gtrsim m_\mu$ the relativistic degrees of freedom in thermal equilibrium are $(\gamma,e,\mu,\nu)$, which correspond to $g_{\ast,S}^{th}=57/4$. For $T\lesssim m_\mu$ the relativistic degrees of freedom in thermal equilibrium are $(\gamma,e,\nu)$, corresponding to $g_{\ast,S}^{th}=43/4$. Imposing conservation of entropy per comoving volume $sa^3$, and $aT_r=$ constant,
\be
\frac{57}{4}(aT_r)^3=\frac{43}{4}(aT_\nu)^3 \quad \Longrightarrow \quad T_\nu=\left( \frac{57}{43}\right)^{1/3}T_r\simeq 1.10 \, T_r \,,
\ee
so that the temperature of the relativistic SM particles in thermal equilibrium $(\gamma,e,\nu)$ is increased with respect to the radion temperature. Let us stress that in this process $a T_\nu$ increases while $a T_r$ remains constant. In this way the contribution of the radion to the effective number of neutrinos is $\Delta N_{\rm eff}=\frac{4}{7}(43/57)^{4/3}\simeq 0.39$, a result which is $\sim 2.4\sigma$ away from the observed value. 
\newline\indent
If the radion decoupling happens at a temperature higher than the mass of several SM particles, when the latter become non-relativistic and annihilate, they heat the temperature of the SM plasma, but not the radion temperature, and again conservation of entropy would lead to $\Delta N_{\rm eff}\ll 1$. The same phenomenon happens if there are phase transitions. As an example we can assume that $T_{r,\FO} > T_{\rm QCD}\simeq 150$ MeV, for which there is also a change in the number of degrees of freedom as quarks condensate into mesons and baryons. Then for $T\gtrsim T_{\rm QCD}$ the relativistic degrees of freedom in thermal equilibrium are: $(\gamma,e,\mu,\nu,u,d,s,g)$ which yields $g_{\ast,S}^{th} = 247/4$. For $T\lesssim T_{\rm QCD}$ the relativistic degrees of freedom in thermal equilibrium are: $(\gamma,e,\mu,\nu,\pi^{\pm,0})$ for which $g_{\ast,S}^{th}=69/4$. The relativistic particles in thermal equilibrium for $T<T_{\rm QCD}$ are heated, with respect to the temperature of the radions, by the amount $(247/69)^{1/3}\simeq 1.53$. When the temperature falls to $T=m_\mu$ muons annihilate and heat the photons. For $T\gtrsim m_\mu$ the degrees of freedom are $(\gamma,e,\mu,\nu)$, which correspond to $g_{\ast,S}^{th}=57/4$, and for $T\lesssim m_\mu$ the degrees of freedom, $(\gamma,e,\nu)$, correspond to $g_{\ast,S}^{th}=43/4$. Finally when the temperature falls to $T=m_e$ electrons annihilate into photons and for $T\gtrsim m_e$ the degrees of freedom are, as in the Standard Cosmology, ($e,\gamma$) with $g_{\ast,S}^{th}=11/2$ and for $T\lesssim m_e$ the only degree of freedom is the photon with $g_{\ast,S}^{th}=2$. Then the heating of photons with respect to the radion and the neutrino is 
\be
T_\gamma=(247/69)^{1/3}(57/43)^{1/3}(11/4)^{1/3} T_r \,, \qquad T_\gamma=(11/4)^{1/3} T_\nu
 \,,  \label{eq:TgammaTradion}
\ee
and identifying the contribution of decoupled neutrinos, and the decoupled radion, to $g_\ast$ as
\be
\Delta g_\ast^{\nu,dec}=\frac{7}{4}\Delta N_{\rm eff}\left(\frac{T_\nu}{T_\gamma} \right)^{4}=g_{\ast}^{r,dec}=\left(\frac{T_r}{T_\gamma} \right)^{4} \Longrightarrow \Delta N_{\rm eff}=\frac{4}{7} \left(\frac{T_r}{T_\nu} \right)^{4}
\label{eq:DeltaNeff}
\ee
we obtain, using Eq.~(\ref{eq:TgammaTradion})
\be
\Delta N_{\rm eff}=\frac{4}{7}\left(\frac{43}{57}\cdot \frac{69}{247}\right)^{4/3}\simeq 0.07  \,,
\ee
 in good agreement with Planck data. 
 \newline\indent
 Finally let us point out that, in this case, there should be a relic background of radions from the time of their decoupling, at a given temperature $T_r$, similar to the CMB radiation at $T_\gamma\simeq 2.73 $ \textbf{K}. The temperature of the radion cosmic background $T_r$ should be smaller than $T_\gamma$, and its actual value does depend on the value of the  decoupling temperature $T_{r,\FO}$. For instance for the case considered above that $T_{r,\FO} \gtrsim \Lambda_{\rm QCD}$ we obtain $T_r\simeq 1.16$  \textbf{K}.

\subsection{The cosmic microwave background bounds}
Free electrons from DM annihilation scatter the cosmic microwave background (CMB) photons and modify the measured anisotropies of the CMB. This leads to upper bounds on the value of the annihilation cross-section $\langle \sigma v\rangle$ for different final states:  $\ell^+\ell^-,q\bar q, \gamma\gamma$~\cite{Planck:2015fie}, the strongest bounds coming from the $e^+e^-$ channel. The observed bounds do depend on the value of the DM mass. For instance for $m_\chi \simeq 1$ (10) GeV the upper bound is $\langle \sigma v\rangle \lesssim 10^{-27}\cm^3/s\ (7\times 10^{-27}\cm^3/s)$. In our model, as we can see from the right panel of Fig.~\ref{fig:Td}, for $m_\chi\simeq 1$ (10) GeV, we get $\langle\sigma_\mu v\rangle \simeq 5\times 10^{-35}\cm^3/s\ (\langle\sigma_b v\rangle \simeq 5\times 10^{-33}\cm^3/s)$, while the cross sections for the $e^+e^-$ channel is $\langle\sigma_e v\rangle=(m_e/m_f)^2\langle\sigma_f v\rangle$, for $f=\mu,b$ respectively, far below the CMB bounds. 

Finally, according to our analysis in Sec.~\ref{sec:indirect}, our model can effortlessly pass the CMB bounds also for the $r \to e^+e^-$ channel.

\subsection{Bounds from cosmic rays}
Measurements of the spectrum of a wide range of cosmic ray species has been performed by the Alpha Magnetic Spectrometer AMS-02 instrument~\cite{AMS:2021nhj}. Current measurements from the antiproton to proton ratio by AMS-02~\cite{Giesen:2015ufa}, provide constraints for different annihilation channels as $\chi\bar\chi\to b \bar b,\, \ell^+\ell^-,\, \gamma\gamma$ which are sensitive to the DM mass. In particular, for $m_\chi\simeq 10$ GeV, they set the 90\% CL bound $\langle\sigma_b v\rangle \lesssim 3\times 10^{-27}\cm^3/s$. As $b\bar b$ is the channel dominating the cross section $\langle \sum_f\sigma_f v\rangle$ for $m_\chi=10$ GeV, the results from the right panel of Fig.~\ref{fig:Td} lead to $\langle\sigma_b v\rangle\simeq 5\times 10^{-33}\cm^3/s$, below the AMS bound. Strongest bounds are derived from the AMS-02 positron to electron+positron ratio~\cite{Bergstrom:2013jra}, even if DM is not responsible for the positron excess, for lepton channels $\ell^+\ell^-$, in particular for the $e^+e^-$ channel for which $\langle\sigma_e v\rangle\lesssim 10^{-28}\cm^3/s$ for $m_\chi\simeq 10$ GeV. In our model the cross sections scale as $\langle\sigma_f v\rangle\propto m_f^2$, so that at the decoupling temperature $T_d$, and for the same value of $m_\chi$, one can rescale the coupling as $\langle \sigma_\ell v\rangle=(m_\ell/m_b)^2\langle\sigma_b v\rangle$ and those bounds are easily evaded.

\subsection{Bounds from the Galactic Center}

The High Energy Stereoscopic System (H.E.S.S.)~\cite{HESS:2022ygk} and Fermi-LAT~\cite{McDaniel:2023bju} collaborations have presented limits on the possible gamma-ray line strength. In particular for the annihilation channel $\chi\bar\chi\to \gamma\gamma$ they find~\cite{HESS:2013rld,Albert:2014hwa} upper bounds on the annihilation cross sections as $\langle\sigma_\gamma v\rangle\lesssim 10^{-29}\cm^3/s$ ($10^{-30}\cm^3/s$) for $m_\chi=1$ (0.1) GeV. On the other hand, from the right panel of Fig.~\ref{fig:Td} one can easily deduce that for $m_\chi\lesssim 10$ GeV, $\langle\sigma_\gamma v\rangle\lesssim 10^{-36}\cm^3/s$. Moreover H.E.S.S.~\cite{HESS:2022ygk}  finds the bound $\langle\sigma_\tau v\rangle\lesssim 10^{-26}\cm^3/s$, for the annihilation channel $\chi\bar\chi\to \tau^+\tau^-$ and any value of $m_\chi >100$ GeV, and Fermi-LAT~\cite{McDaniel:2023bju} provides the bound $\langle\sigma_\tau v\rangle\lesssim 2\times 10^{-27}\cm^3/s$ for the same channel $\chi\bar\chi\to \tau^+\tau^-$ and $m_\chi\simeq 2$ GeV, while the right panel plot of Fig.~\ref{fig:Td} provides in our model the upper bound $\langle\sigma_\tau v\rangle\simeq 6\times 10^{-34}\cm^3/s$, orders of magnitude below.

\section{Summary and outlook}
\label{sec:conclusions}

In this paper we have considered a 5D model with three 4D branes along the extra dimension: the UV brane at the Planck scale, the TeV brane, and the IR brane, or dark brane, at a scale around the GeV scale.
\begin{itemize}
\item
The IR brane triggers a confinement/deconfinement first order phase transition which produces a stochastic gravitational waves background that can accommodate the recent signal observed by the PTA experiments, an issue studied in detail in a previous publication~\cite{Megias:2023kiy}.
\item
The SM is localized on the TeV brane, thus solving the hierarchy problem, the warp factor relating the Planck and TeV scales, as in the original RS proposal.
\item
The dark sector, and in particular the dark matter, considered as a Dirac fermion, is localized on the IR brane.
\item
The KK gravitons and the radion  are bulk propagating fields that connect the TeV and the IR branes. However the interaction of gravitons with the TeV brane is much weaker than the interaction of radions, so we have considered only radions as mediators between the TeV and IR branes. 
\end{itemize}

\begin{figure}[t]
\centering
\includegraphics[width=8.5cm]{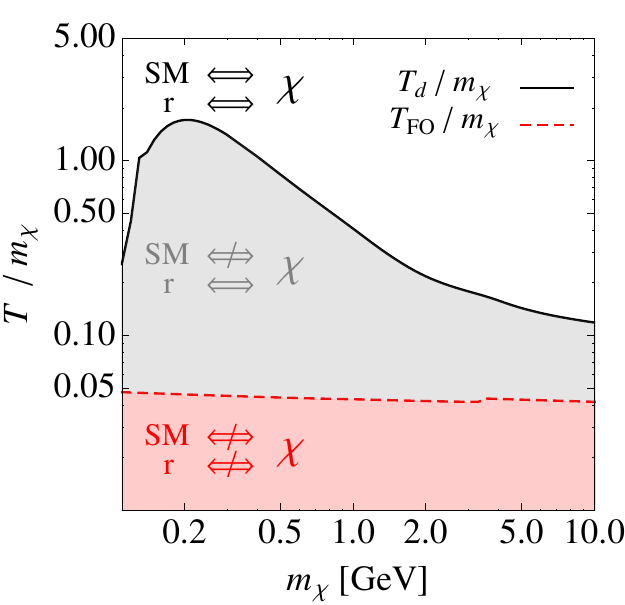} 
\caption{\it The temperatures $T_d/m_\chi$ (solid line) and $T_{\FO}/m_\chi$ (dashed line) as functions of $m_\chi$. The symbol $\Longleftrightarrow$ ($\slashed{\Longleftrightarrow}$) between two species means that they are (not) in thermal equilibrium.}
\label{fig:TdTFO}
\end{figure}

The dark matter interacts with the SM via radion mediation and can annihilate into SM fermions ($f$) as well as massless gauge bosons, photons ($\gamma$) and gluons ($g$), and into a pair of radions ($r$), as we are considering the region where $m_r<m_\chi$. Then the following thermal history appears for the system SM+DM+radion:

\begin{itemize}
\item
For temperatures $T>T_d$, where $T_d\simeq 0.2\, \textrm{GeV}+0.1\,m_\chi$, the DM is in thermal equilibrium with the SM and the radion sector, via the reactions $\bar\chi\chi\leftrightarrow \bar f f, gg, \gamma\gamma$ and $\chi\bar\chi\leftrightarrow rr$, respectively.  Upper (white) region in Fig.~\ref{fig:TdTFO}. We have also defined the freeze-in temperature for radions with respect to the SM plasma such that, for temperatures $T$ such that $T_{\rm FI}>T>T_d$ the radions are in equilibrium with the SM plasma mainly via the decay and inverse decay reactions $r\leftrightarrow f\bar f,\gamma\gamma,gg$.

\item
At $T\simeq T_d$ the DM decouples, via the reactions $\bar\chi\chi\leftrightarrow \bar f f, gg, \gamma\gamma$, from the SM and keeps thermal equilibrium with radions through the reaction $\bar\chi\chi\to rr$.
For $T_d>T>T_{\FO}=m_\chi/x_{\FO}$, where $x_{\FO}\simeq 24+ \log(m_\chi/\textrm{GeV})$, the DM is in thermal equilibrium with radions via the reaction 
$\bar\chi\chi\to rr$. Middle (gray) region in Fig.~\ref{fig:TdTFO}.  
\item
At $T=T_{\FO}$ the DM goes out of thermal equilibrium and the relic abundance is generated.
\item
For $T<T_{\FO}$ the DM and the radion are out of thermal equilibrium. Lower (red) region in Fig.~\ref{fig:TdTFO}. Still the radion keeps thermal equilibrium with the SM plasma via the decay and inverse decays. Its abundance then goes to zero non-relativistically, for $T<m_r$, as $e^{-m_r/T}$.

\end{itemize} 

The temperatures $T_d$ (solid line) and $T_{\FO}$ (dashed line) are shown in Fig.~\ref{fig:TdTFO} as functions of $m_\chi$, where the value of $T_{\FO}$ already assumes that $\Omega_\chi h^2=0.12$, according with the cosmological observations. The temperature $T_{\rm FI}\gtrsim T_d$ at which the radion enters thermal equilibrium with the SM plasma can be easily deduced from Fig.~\ref{fig:sigma}.

In the calculation of the relic density we have assumed that $m_r<m_\chi$ so that, neglecting the value of~$m_r$, the value of $\Omega_\chi h^2$ mainly depends on $m_\chi$ and $\trho_1$. As the relic density incorporates the interaction of radion with the DM, localized in the IR brane, while direct measurements are controlled by the interaction of the radion with both the DM and with the SM, it is very easy to avoid bounds from direct measurements of DM scattering on nucleon and electrons while keeping the relic density safe. In fact, for a given value of the DM mass $m_\chi$, direct measurements translate into lower bounds on the radion mass $m_r$, while for $m_\chi\gtrsim 0.5$~GeV and data from DM-nucleon scattering, the neutrino floor translates into an upper bound on the value $m_r$. This gives an allowed window on the radion mass, well in agreement with the assumption  $m_r<m_\chi$ and consistent with the observed relic density.

In the same way indirect measurements, based on the annihilation processes $\bar\chi\chi\to \textrm{SM} + \textrm{SM}$, are easily avoided as the messenger radion is very weakly coupled with the TeV brane, where the SM is localized. We have found that bounds from radion cosmology and BBN, after imposing that their lifetime is smaller than 10 sec, and the constraint that radions do not perturb the DM relic density lead to the bound $m_r\gtrsim 1$ MeV on the radion masses.
Moreover, for $m_r<1$ MeV the theory is excluded by the BBN conditions, unless $m_r\lesssim$ 10 keV, in which case radions decay with a lifetime greater than the age of the universe, thus behave as stable particle, and could be considered as a DM ingredient. This case, where radions are relativistic at the BBN time, is constrained by Planck bounds on $\Delta N_{\rm eff}$ which, in turn, depends on the temperature $T_{r,\FO}$ at which radions freeze-out and decouple from the SM plasma.

For the same reason, bounds from accelerator searches are easily avoided. Notice here that many of these searches are based on the mediator decaying invisibly. In our case the radion with a mass $m_r<m_\chi$ cannot decay into $\bar\chi\chi$ as this channel is kinematically forbidden. However it can decay into SM fields with a decay width such that $\Gamma_r/m_r\ll 1$ and a macroscopic lifetime $\tau_r$. Depending on the values of $\trho_1$ and $m_\chi$, the radion can decay inside the detector, or outside it, in which case the bounds for invisible decay should apply. 

As it was pointed out in the introduction, the work of Ref.~\cite{Ferrante:2023bcz} is also considering a three-brane RS model putting the (scalar) DM on the dark GeV brane. Even though this is a very similar set-up to our own, we share some important and crucial differences worth to be extensively mentioned here.
In particular, some relevant differences between our scenario and that in Ref.~\cite{Ferrante:2023bcz} are: \textit{i)}~Unlike our construction, Ref.~\cite{Ferrante:2023bcz} considers the case where $m_r>m_S$ so that the relic density through radion production is via the mechanism of forbidden DM, which requires some degree of tuning between the radion and DM masses. \textit{ii)}~In our model the SM fermions are in the TeV brane, so that the flavor problem is the same as in the SM, while in Ref.~\cite{Ferrante:2023bcz} only the Higgs and $t_R$ are in the TeV brane, while the rest of fermions, including the $(t,b)_L$ doublet, are (elementary) in the Planck brane, making it difficult to reproduce the top mass. Moreover in our model as light quarks are in the TeV brane their coupling to the mediator radion is moderate leaving a window for direct detection in the future outside the neutrino floor. \textit{iii)}~Lastly, for values of the DM mass much smaller than the dark brane scale there should appear some little hierarchy problem, similar to what would appear in RS theories for an extremely light Higgs.

Finally, in this paper we have considered the simplest case where only DM, a Dirac fermion, is living in the IR brane. Other scenarios are worth exploring, as \textit{e.g.}~assuming that DM is a real scalar, a Majorana fermion, or a vector boson. On the other hand a more complicated dark sector can be localized in the IR brane, including dark photons and a dark Higgs, while the DM can get its mass from some spontaneous breaking mechanism where the dark Higgs gets a VEV and gives a mass to the dark photons. Finally our 5D model has a holographic dual interpretation, which can be understood as a conformal 4D theory with some mechanism of spontaneous conformal symmetry breaking, which can be clearly exploited in forthcoming analyses.

\begin{acknowledgments}
  \noindent

We thank A.~Donini, M.~G.~Folgado, J.~Herrero, G.~Landini, A.~Mu\~{n}oz and N.~Rius for useful comments on the radion couplings in an early version of this work. FK would like to thank Ayuki Kamada for helpful discussions about direct and indirect Dark Matter searches. The work of EM is supported by the project PID2020-114767GB-I00 and by the Ram\'on y Cajal Program under Grant RYC-2016-20678 funded by MCIN/AEI/10.13039/501100011033 and by ``FSE Investing in your future'', by Junta de Andaluc\'{\i}a under Grant FQM-225, and  by the ``Pr\'orrogas de Contratos Ram\'on y Cajal'' Program of the University of Granada. The work of SP was supported by National Science Centre, Poland, grant DEC-2018/31/B/ST2/02283. The work of MQ is supported by the Departament d'Empresa i Coneixement, Generalitat de Catalunya, Grant No.~2021 SGR 00649, and by the Ministerio de Econom\'{\i}a y Competitividad, Grant No.~PID2020-115845GB-I00. IFAE is partially funded by the CERCA program of the Generalitat de Catalunya.
  
\end{acknowledgments}

\bibliographystyle{JHEP}
\bibliography{refs}

\providecommand{\href}[2]{#2}\begingroup\raggedright\begin{thebibliography}{10}

\bibitem{ParticleDataGroup:2022pth}
{\scshape Particle Data Group} collaboration, R.~L. Workman et~al.,
  \emph{{Review of Particle Physics}},
  \href{http://dx.doi.org/10.1093/ptep/ptac097}{\emph{PTEP} {\bfseries 2022}
  (2022) 083C01}.

\bibitem{Randall:1999ee}
L.~Randall and R.~Sundrum, \emph{{A Large mass hierarchy from a small extra
  dimension}}, \href{http://dx.doi.org/10.1103/PhysRevLett.83.3370}{\emph{Phys.
  Rev. Lett.} {\bfseries 83} (1999) 3370--3373},
  [\href{https://arxiv.org/abs/hep-ph/9905221}{{\ttfamily hep-ph/9905221}}].

\bibitem{Goldberger:1999uk}
W.~D. Goldberger and M.~B. Wise, \emph{{Modulus stabilization with bulk
  fields}}, \href{http://dx.doi.org/10.1103/PhysRevLett.83.4922}{\emph{Phys.
  Rev. Lett.} {\bfseries 83} (1999) 4922--4925},
  [\href{https://arxiv.org/abs/hep-ph/9907447}{{\ttfamily hep-ph/9907447}}].

\bibitem{Lee:2021wau}
S.~J. Lee, Y.~Nakai and M.~Suzuki, \emph{{Multiple hierarchies from a warped
  extra dimension}},
  \href{http://dx.doi.org/10.1007/JHEP02(2022)050}{\emph{JHEP} {\bfseries 02}
  (2022) 050}, [\href{https://arxiv.org/abs/2109.10938}{{\ttfamily
  2109.10938}}].

\bibitem{Girmohanta:2023sjv}
S.~Girmohanta, S.~J. Lee, Y.~Nakai and M.~Suzuki, \emph{{Multi-brane
  cosmology}}, \href{http://dx.doi.org/10.1007/JHEP07(2023)182}{\emph{JHEP}
  {\bfseries 07} (2023) 182},
  [\href{https://arxiv.org/abs/2304.05586}{{\ttfamily 2304.05586}}].

\bibitem{Simon_2019}
J.~D. Simon, \emph{The faintest dwarf galaxies},
  \href{http://dx.doi.org/10.1146/annurev-astro-091918-104453}{\emph{Annual
  Review of Astronomy and Astrophysics} {\bfseries 57} (Aug., 2019) 375–415}.

\bibitem{Salucci_2019}
P.~Salucci, \emph{The distribution dark matter in galaxies},
  \href{http://dx.doi.org/10.1007/s00159-018-0113-1}{\emph{The Astronomy and
  Astrophysics Review} {\bfseries 27} (Feb., 2019) }.

\bibitem{Allen_2011}
S.~W. Allen, A.~E. Evrard and A.~B. Mantz, \emph{Cosmological parameters from
  observations of galaxy clusters},
  \href{http://dx.doi.org/10.1146/annurev-astro-081710-102514}{\emph{Annual
  Review of Astronomy and Astrophysics} {\bfseries 49} (Sept., 2011)
  409–470}.

\bibitem{Jungman:1995df}
G.~Jungman, M.~Kamionkowski and K.~Griest, \emph{{Supersymmetric dark matter}},
  \href{http://dx.doi.org/10.1016/0370-1573(95)00058-5}{\emph{Phys. Rept.}
  {\bfseries 267} (1996) 195--373},
  [\href{https://arxiv.org/abs/hep-ph/9506380}{{\ttfamily hep-ph/9506380}}].

\bibitem{tHooft:1979rat}
G.~'t~Hooft, \emph{{Naturalness, chiral symmetry, and spontaneous chiral
  symmetry breaking}},
  \href{http://dx.doi.org/10.1007/978-1-4684-7571-5_9}{\emph{NATO Sci. Ser. B}
  {\bfseries 59} (1980) 135--157}.

\bibitem{Xu:2023wog}
H.~Xu et~al., \emph{{Searching for the Nano-Hertz Stochastic Gravitational Wave
  Background with the Chinese Pulsar Timing Array Data Release I}},
  \href{http://dx.doi.org/10.1088/1674-4527/acdfa5}{\emph{Res. Astron.
  Astrophys.} {\bfseries 23} (2023) 075024},
  [\href{https://arxiv.org/abs/2306.16216}{{\ttfamily 2306.16216}}].

\bibitem{Reardon:2023gzh}
D.~J. Reardon et~al., \emph{{Search for an Isotropic Gravitational-wave
  Background with the Parkes Pulsar Timing Array}},
  \href{http://dx.doi.org/10.3847/2041-8213/acdd02}{\emph{Astrophys. J. Lett.}
  {\bfseries 951} (2023) L6},
  [\href{https://arxiv.org/abs/2306.16215}{{\ttfamily 2306.16215}}].

\bibitem{EPTA:2023fyk}
{\scshape EPTA, InPTA:} collaboration, J.~Antoniadis et~al., \emph{{The second
  data release from the European Pulsar Timing Array - III. Search for
  gravitational wave signals}},
  \href{http://dx.doi.org/10.1051/0004-6361/202346844}{\emph{Astron.
  Astrophys.} {\bfseries 678} (2023) A50},
  [\href{https://arxiv.org/abs/2306.16214}{{\ttfamily 2306.16214}}].

\bibitem{NANOGrav:2023gor}
{\scshape NANOGrav} collaboration, G.~Agazie et~al., \emph{{The NANOGrav 15 yr
  Data Set: Evidence for a Gravitational-wave Background}},
  \href{http://dx.doi.org/10.3847/2041-8213/acdac6}{\emph{Astrophys. J. Lett.}
  {\bfseries 951} (2023) L8},
  [\href{https://arxiv.org/abs/2306.16213}{{\ttfamily 2306.16213}}].

\bibitem{Megias:2023kiy}
E.~Meg\'\i{}as, G.~Nardini and M.~Quir\'os, \emph{{Pulsar timing array
  stochastic background from light Kaluza-Klein resonances}},
  \href{http://dx.doi.org/10.1103/PhysRevD.108.095017}{\emph{Phys. Rev. D}
  {\bfseries 108} (2023) 095017},
  [\href{https://arxiv.org/abs/2306.17071}{{\ttfamily 2306.17071}}].

\bibitem{Lee:2013bua}
H.~M. Lee, M.~Park and V.~Sanz, \emph{{Gravity-mediated (or Composite) Dark
  Matter}}, \href{http://dx.doi.org/10.1140/epjc/s10052-014-2715-8}{\emph{Eur.
  Phys. J. C} {\bfseries 74} (2014) 2715},
  [\href{https://arxiv.org/abs/1306.4107}{{\ttfamily 1306.4107}}].

\bibitem{Folgado:2019sgz}
M.~G. Folgado, A.~Donini and N.~Rius, \emph{{Gravity-mediated Scalar Dark
  Matter in Warped Extra-Dimensions}},
  \href{http://dx.doi.org/10.1007/JHEP01(2020)161}{\emph{JHEP} {\bfseries 01}
  (2020) 161}, [\href{https://arxiv.org/abs/1907.04340}{{\ttfamily
  1907.04340}}].

\bibitem{deGiorgi:2021xvm}
A.~de~Giorgi and S.~Vogl, \emph{{Dark matter interacting via a massive spin-2
  mediator in warped extra-dimensions}},
  \href{http://dx.doi.org/10.1007/JHEP11(2021)036}{\emph{JHEP} {\bfseries 11}
  (2021) 036}, [\href{https://arxiv.org/abs/2105.06794}{{\ttfamily
  2105.06794}}].

\bibitem{deGiorgi:2022yha}
A.~de~Giorgi and S.~Vogl, \emph{{Warm dark matter from a gravitational
  freeze-in in extra dimensions}},
  \href{http://dx.doi.org/10.1007/JHEP04(2023)032}{\emph{JHEP} {\bfseries 04}
  (2023) 032}, [\href{https://arxiv.org/abs/2208.03153}{{\ttfamily
  2208.03153}}].

\bibitem{vonHarling:2012sz}
B.~von Harling and K.~L. McDonald, \emph{{Secluded Dark Matter Coupled to a
  Hidden CFT}}, \href{http://dx.doi.org/10.1007/JHEP08(2012)048}{\emph{JHEP}
  {\bfseries 08} (2012) 048},
  [\href{https://arxiv.org/abs/1203.6646}{{\ttfamily 1203.6646}}].

\bibitem{Ferrante:2023bcz}
S.~Ferrante, A.~Ismail, S.~J. Lee and Y.~Lee, \emph{{Forbidden conformal dark
  matter at a GeV}},
  \href{http://dx.doi.org/10.1007/JHEP11(2023)186}{\emph{JHEP} {\bfseries 11}
  (2023) 186}, [\href{https://arxiv.org/abs/2308.16219}{{\ttfamily
  2308.16219}}].

\bibitem{Cai:2021mrw}
H.~Cai, \emph{{Radion dynamics in the multibrane Randall-Sundrum model}},
  \href{http://dx.doi.org/10.1103/PhysRevD.105.075009}{\emph{Phys. Rev. D}
  {\bfseries 105} (2022) 075009},
  [\href{https://arxiv.org/abs/2109.09681}{{\ttfamily 2109.09681}}].

\bibitem{Foot:2014uba}
R.~Foot and S.~Vagnozzi, \emph{{Dissipative hidden sector dark matter}},
  \href{http://dx.doi.org/10.1103/PhysRevD.91.023512}{\emph{Phys. Rev. D}
  {\bfseries 91} (2015) 023512},
  [\href{https://arxiv.org/abs/1409.7174}{{\ttfamily 1409.7174}}].

\bibitem{Breitbach:2018ddu}
M.~Breitbach, J.~Kopp, E.~Madge, T.~Opferkuch and P.~Schwaller, \emph{{Dark,
  Cold, and Noisy: Constraining Secluded Hidden Sectors with Gravitational
  Waves}}, \href{http://dx.doi.org/10.1088/1475-7516/2019/07/007}{\emph{JCAP}
  {\bfseries 07} (2019) 007},
  [\href{https://arxiv.org/abs/1811.11175}{{\ttfamily 1811.11175}}].

\bibitem{Fairbairn:2019xog}
M.~Fairbairn, E.~Hardy and A.~Wickens, \emph{{Hearing without seeing:
  gravitational waves from hot and cold hidden sectors}},
  \href{http://dx.doi.org/10.1007/JHEP07(2019)044}{\emph{JHEP} {\bfseries 07}
  (2019) 044}, [\href{https://arxiv.org/abs/1901.11038}{{\ttfamily
  1901.11038}}].

\bibitem{Brax:2019koq}
P.~Brax, S.~Fichet and P.~Tanedo, \emph{{The Warped Dark Sector}},
  \href{http://dx.doi.org/10.1016/j.physletb.2019.135012}{\emph{Phys. Lett. B}
  {\bfseries 798} (2019) 135012},
  [\href{https://arxiv.org/abs/1906.02199}{{\ttfamily 1906.02199}}].

\bibitem{Bento:2023flt}
M.~P. Bento, H.~E. Haber and J.~a.~P. Silva, \emph{{Classes of complete dark
  photon models constrained by Z-physics}},
  \href{http://dx.doi.org/10.1016/j.physletb.2024.138501}{\emph{Phys. Lett. B}
  {\bfseries 850} (2024) 138501},
  [\href{https://arxiv.org/abs/2311.04976}{{\ttfamily 2311.04976}}].

\bibitem{DeWolfe:1999cp}
O.~DeWolfe, D.~Z. Freedman, S.~S. Gubser and A.~Karch, \emph{{Modeling the
  fifth-dimension with scalars and gravity}},
  \href{http://dx.doi.org/10.1103/PhysRevD.62.046008}{\emph{Phys. Rev. D}
  {\bfseries 62} (2000) 046008},
  [\href{https://arxiv.org/abs/hep-th/9909134}{{\ttfamily hep-th/9909134}}].

\bibitem{Batell:2008me}
B.~Batell, T.~Gherghetta and D.~Sword, \emph{{The Soft-Wall Standard Model}},
  \href{http://dx.doi.org/10.1103/PhysRevD.78.116011}{\emph{Phys. Rev. D}
  {\bfseries 78} (2008) 116011},
  [\href{https://arxiv.org/abs/0808.3977}{{\ttfamily 0808.3977}}].

\bibitem{Cabrer:2009we}
J.~A. Cabrer, G.~von Gersdorff and M.~Quir\'os, \emph{{Soft-Wall
  Stabilization}},
  \href{http://dx.doi.org/10.1088/1367-2630/12/7/075012}{\emph{New J. Phys.}
  {\bfseries 12} (2010) 075012},
  [\href{https://arxiv.org/abs/0907.5361}{{\ttfamily 0907.5361}}].

\bibitem{Cabrer:2010si}
J.~A. Cabrer, G.~von Gersdorff and M.~Quir\'os, \emph{{Warped Electroweak
  Breaking Without Custodial Symmetry}},
  \href{http://dx.doi.org/10.1016/j.physletb.2011.01.058}{\emph{Phys. Lett. B}
  {\bfseries 697} (2011) 208--214},
  [\href{https://arxiv.org/abs/1011.2205}{{\ttfamily 1011.2205}}].

\bibitem{Cabrer:2011fb}
J.~A. Cabrer, G.~von Gersdorff and M.~Quir\'os, \emph{{Suppressing Electroweak
  Precision Observables in 5D Warped Models}},
  \href{http://dx.doi.org/10.1007/JHEP05(2011)083}{\emph{JHEP} {\bfseries 05}
  (2011) 083}, [\href{https://arxiv.org/abs/1103.1388}{{\ttfamily 1103.1388}}].

\bibitem{Cabrer:2011vu}
J.~A. Cabrer, G.~von Gersdorff and M.~Quir\'os, \emph{{Improving Naturalness in
  Warped Models with a Heavy Bulk Higgs Boson}},
  \href{http://dx.doi.org/10.1103/PhysRevD.84.035024}{\emph{Phys. Rev. D}
  {\bfseries 84} (2011) 035024},
  [\href{https://arxiv.org/abs/1104.3149}{{\ttfamily 1104.3149}}].

\bibitem{Cabrer:2011qb}
J.~A. Cabrer, G.~von Gersdorff and M.~Quir\'os, \emph{{Flavor Phenomenology in
  General 5D Warped Spaces}},
  \href{http://dx.doi.org/10.1007/JHEP01(2012)033}{\emph{JHEP} {\bfseries 01}
  (2012) 033}, [\href{https://arxiv.org/abs/1110.3324}{{\ttfamily 1110.3324}}].

\bibitem{Megias:2020vek}
E.~Meg\'\i{}as, G.~Nardini and M.~Quir\'os, \emph{{Gravitational Imprints from
  Heavy Kaluza-Klein Resonances}},
  \href{http://dx.doi.org/10.1103/PhysRevD.102.055004}{\emph{Phys. Rev. D}
  {\bfseries 102} (2020) 055004},
  [\href{https://arxiv.org/abs/2005.04127}{{\ttfamily 2005.04127}}].

\bibitem{Davoudiasl:1999jd}
H.~Davoudiasl, J.~L. Hewett and T.~G. Rizzo, \emph{{Phenomenology of the
  Randall-Sundrum Gauge Hierarchy Model}},
  \href{http://dx.doi.org/10.1103/PhysRevLett.84.2080}{\emph{Phys. Rev. Lett.}
  {\bfseries 84} (2000) 2080},
  [\href{https://arxiv.org/abs/hep-ph/9909255}{{\ttfamily hep-ph/9909255}}].

\bibitem{Contino:2001nj}
R.~Contino, L.~Pilo, R.~Rattazzi and A.~Strumia, \emph{{Graviton loops and
  brane observables}},
  \href{http://dx.doi.org/10.1088/1126-6708/2001/06/005}{\emph{JHEP} {\bfseries
  06} (2001) 005}, [\href{https://arxiv.org/abs/hep-ph/0103104}{{\ttfamily
  hep-ph/0103104}}].

\bibitem{Csaki:2000zn}
C.~Csaki, M.~L. Graesser and G.~D. Kribs, \emph{{Radion dynamics and
  electroweak physics}},
  \href{http://dx.doi.org/10.1103/PhysRevD.63.065002}{\emph{Phys. Rev. D}
  {\bfseries 63} (2001) 065002},
  [\href{https://arxiv.org/abs/hep-th/0008151}{{\ttfamily hep-th/0008151}}].

\bibitem{Csaki:2007ns}
C.~Csaki, J.~Hubisz and S.~J. Lee, \emph{{Radion phenomenology in realistic
  warped space models}},
  \href{http://dx.doi.org/10.1103/PhysRevD.76.125015}{\emph{Phys. Rev. D}
  {\bfseries 76} (2007) 125015},
  [\href{https://arxiv.org/abs/0705.3844}{{\ttfamily 0705.3844}}].

\bibitem{Konstandin:2010cd}
T.~Konstandin, G.~Nardini and M.~Quir\'os, \emph{{Gravitational Backreaction
  Effects on the Holographic Phase Transition}},
  \href{http://dx.doi.org/10.1103/PhysRevD.82.083513}{\emph{Phys. Rev. D}
  {\bfseries 82} (2010) 083513},
  [\href{https://arxiv.org/abs/1007.1468}{{\ttfamily 1007.1468}}].

\bibitem{Hooper:2018kfv}
D.~Hooper, \emph{{TASI Lectures on Indirect Searches For Dark Matter}},
  {\emph{PoS} {\bfseries TASI2018} (2019) 010},
  [\href{https://arxiv.org/abs/1812.02029}{{\ttfamily 1812.02029}}].

\bibitem{Gondolo:1990dk}
P.~Gondolo and G.~Gelmini, \emph{{Cosmic abundances of stable particles:
  Improved analysis}},
  \href{http://dx.doi.org/10.1016/0550-3213(91)90438-4}{\emph{Nucl. Phys. B}
  {\bfseries 360} (1991) 145--179}.

\bibitem{Giudice:2000av}
G.~F. Giudice, R.~Rattazzi and J.~D. Wells, \emph{{Graviscalars from higher
  dimensional metrics and curvature Higgs mixing}},
  \href{http://dx.doi.org/10.1016/S0550-3213(00)00686-6}{\emph{Nucl. Phys. B}
  {\bfseries 595} (2001) 250--276},
  [\href{https://arxiv.org/abs/hep-ph/0002178}{{\ttfamily hep-ph/0002178}}].

\bibitem{Feng:2008ya}
J.~L. Feng and J.~Kumar, \emph{{The WIMPless Miracle: Dark-Matter Particles
  without Weak-Scale Masses or Weak Interactions}},
  \href{http://dx.doi.org/10.1103/PhysRevLett.101.231301}{\emph{Phys. Rev.
  Lett.} {\bfseries 101} (2008) 231301},
  [\href{https://arxiv.org/abs/0803.4196}{{\ttfamily 0803.4196}}].

\bibitem{DAgnolo:2015ujb}
R.~T. D'Agnolo and J.~T. Ruderman, \emph{{Light Dark Matter from Forbidden
  Channels}},
  \href{http://dx.doi.org/10.1103/PhysRevLett.115.061301}{\emph{Phys. Rev.
  Lett.} {\bfseries 115} (2015) 061301},
  [\href{https://arxiv.org/abs/1505.07107}{{\ttfamily 1505.07107}}].

\bibitem{Agrawal:2010fh}
P.~Agrawal, Z.~Chacko, C.~Kilic and R.~K. Mishra, \emph{{A Classification of
  Dark Matter Candidates with Primarily Spin-Dependent Interactions with
  Matter}},  \href{https://arxiv.org/abs/1003.1912}{{\ttfamily 1003.1912}}.

\bibitem{Ellis:2018dmb}
J.~Ellis, N.~Nagata and K.~A. Olive, \emph{{Uncertainties in WIMP Dark Matter
  Scattering Revisited}},
  \href{http://dx.doi.org/10.1140/epjc/s10052-018-6047-y}{\emph{Eur. Phys. J.
  C} {\bfseries 78} (2018) 569},
  [\href{https://arxiv.org/abs/1805.09795}{{\ttfamily 1805.09795}}].

\bibitem{XENON:2019zpr}
{\scshape XENON} collaboration, E.~Aprile et~al., \emph{{Search for Light Dark
  Matter Interactions Enhanced by the Migdal Effect or Bremsstrahlung in
  XENON1T}},
  \href{http://dx.doi.org/10.1103/PhysRevLett.123.241803}{\emph{Phys. Rev.
  Lett.} {\bfseries 123} (2019) 241803},
  [\href{https://arxiv.org/abs/1907.12771}{{\ttfamily 1907.12771}}].

\bibitem{XENON:2019gfn}
{\scshape XENON} collaboration, E.~Aprile et~al., \emph{{Light Dark Matter
  Search with Ionization Signals in XENON1T}},
  \href{http://dx.doi.org/10.1103/PhysRevLett.123.251801}{\emph{Phys. Rev.
  Lett.} {\bfseries 123} (2019) 251801},
  [\href{https://arxiv.org/abs/1907.11485}{{\ttfamily 1907.11485}}].

\bibitem{DarkSide-50:2022qzh}
{\scshape DarkSide-50} collaboration, P.~Agnes et~al., \emph{{Search for
  low-mass dark matter WIMPs with 12~ton-day exposure of DarkSide-50}},
  \href{http://dx.doi.org/10.1103/PhysRevD.107.063001}{\emph{Phys. Rev. D}
  {\bfseries 107} (2023) 063001},
  [\href{https://arxiv.org/abs/2207.11966}{{\ttfamily 2207.11966}}].

\bibitem{CRESST:2019jnq}
{\scshape CRESST} collaboration, A.~H. Abdelhameed et~al., \emph{{First results
  from the CRESST-III low-mass dark matter program}},
  \href{http://dx.doi.org/10.1103/PhysRevD.100.102002}{\emph{Phys. Rev. D}
  {\bfseries 100} (2019) 102002},
  [\href{https://arxiv.org/abs/1904.00498}{{\ttfamily 1904.00498}}].

\bibitem{Essig:2015cda}
R.~Essig, M.~Fernandez-Serra, J.~Mardon, A.~Soto, T.~Volansky and T.-T. Yu,
  \emph{{Direct Detection of sub-GeV Dark Matter with Semiconductor Targets}},
  \href{http://dx.doi.org/10.1007/JHEP05(2016)046}{\emph{JHEP} {\bfseries 05}
  (2016) 046}, [\href{https://arxiv.org/abs/1509.01598}{{\ttfamily
  1509.01598}}].

\bibitem{XENON:2021qze}
{\scshape XENON} collaboration, E.~Aprile et~al., \emph{{Emission of single and
  few electrons in XENON1T and limits on light dark matter}},
  \href{http://dx.doi.org/10.1103/PhysRevD.106.022001}{\emph{Phys. Rev. D}
  {\bfseries 106} (2022) 022001},
  [\href{https://arxiv.org/abs/2112.12116}{{\ttfamily 2112.12116}}].

\bibitem{DarkSide:2022knj}
{\scshape DarkSide} collaboration, P.~Agnes et~al., \emph{{Search for Dark
  Matter Particle Interactions with Electron Final States with DarkSide-50}},
  \href{http://dx.doi.org/10.1103/PhysRevLett.130.101002}{\emph{Phys. Rev.
  Lett.} {\bfseries 130} (2023) 101002},
  [\href{https://arxiv.org/abs/2207.11968}{{\ttfamily 2207.11968}}].

\bibitem{SENSEI:2023zdf}
{\scshape SENSEI} collaboration, P.~Adari et~al., \emph{{SENSEI: First
  Direct-Detection Results on sub-GeV Dark Matter from SENSEI at SNOLAB}},
  \href{https://arxiv.org/abs/2312.13342}{{\ttfamily 2312.13342}}.

\bibitem{Carpenter:2012rg}
L.~M. Carpenter, A.~Nelson, C.~Shimmin, T.~M.~P. Tait and D.~Whiteson,
  \emph{{Collider searches for dark matter in events with a Z boson and missing
  energy}}, \href{http://dx.doi.org/10.1103/PhysRevD.87.074005}{\emph{Phys.
  Rev. D} {\bfseries 87} (2013) 074005},
  [\href{https://arxiv.org/abs/1212.3352}{{\ttfamily 1212.3352}}].

\bibitem{ATLAS:2012bra}
{\scshape ATLAS} collaboration, G.~Aad et~al., \emph{{Measurement of $ZZ$
  production in $pp$ collisions at $\sqrt{s}=7$ TeV and limits on anomalous
  $ZZZ$ and $ZZ\gamma$ couplings with the ATLAS detector}},
  \href{http://dx.doi.org/10.1007/JHEP03(2013)128}{\emph{JHEP} {\bfseries 03}
  (2013) 128}, [\href{https://arxiv.org/abs/1211.6096}{{\ttfamily 1211.6096}}].

\bibitem{NA64:2023wbi}
{\scshape NA64} collaboration, Y.~M. Andreev et~al., \emph{{Search for Light
  Dark Matter with NA64 at CERN}},
  \href{http://dx.doi.org/10.1103/PhysRevLett.131.161801}{\emph{Phys. Rev.
  Lett.} {\bfseries 131} (2023) 161801},
  [\href{https://arxiv.org/abs/2307.02404}{{\ttfamily 2307.02404}}].

\bibitem{Berlin:2018bsc}
A.~Berlin, N.~Blinov, G.~Krnjaic, P.~Schuster and N.~Toro, \emph{{Dark Matter,
  Millicharges, Axion and Scalar Particles, Gauge Bosons, and Other New Physics
  with LDMX}}, \href{http://dx.doi.org/10.1103/PhysRevD.99.075001}{\emph{Phys.
  Rev. D} {\bfseries 99} (2019) 075001},
  [\href{https://arxiv.org/abs/1807.01730}{{\ttfamily 1807.01730}}].

\bibitem{NA64:2021xzo}
{\scshape NA64} collaboration, Y.~M. Andreev et~al., \emph{{Constraints on New
  Physics in Electron $g-2$ from a Search for Invisible Decays of a Scalar,
  Pseudoscalar, Vector, and Axial Vector}},
  \href{http://dx.doi.org/10.1103/PhysRevLett.126.211802}{\emph{Phys. Rev.
  Lett.} {\bfseries 126} (2021) 211802},
  [\href{https://arxiv.org/abs/2102.01885}{{\ttfamily 2102.01885}}].

\bibitem{Joyce:1997uy}
M.~Joyce and M.~E. Shaposhnikov, \emph{{Primordial magnetic fields,
  right-handed electrons, and the Abelian anomaly}},
  \href{http://dx.doi.org/10.1103/PhysRevLett.79.1193}{\emph{Phys. Rev. Lett.}
  {\bfseries 79} (1997) 1193--1196},
  [\href{https://arxiv.org/abs/astro-ph/9703005}{{\ttfamily
  astro-ph/9703005}}].

\bibitem{Millea:2015qra}
M.~Millea, L.~Knox and B.~Fields, \emph{{New Bounds for Axions and Axion-Like
  Particles with keV-GeV Masses}},
  \href{http://dx.doi.org/10.1103/PhysRevD.92.023010}{\emph{Phys. Rev. D}
  {\bfseries 92} (2015) 023010},
  [\href{https://arxiv.org/abs/1501.04097}{{\ttfamily 1501.04097}}].

\bibitem{Abu-Ajamieh:2017khi}
F.~Abu-Ajamieh, J.~S. Lee and J.~Terning, \emph{{The Light Radion Window}},
  \href{http://dx.doi.org/10.1007/JHEP10(2018)050}{\emph{JHEP} {\bfseries 10}
  (2018) 050}, [\href{https://arxiv.org/abs/1711.02697}{{\ttfamily
  1711.02697}}].

\bibitem{Abu-Ajamieh:2018brk}
F.~Abu-Ajamieh, \emph{{The Radion as a Dark Matter Candidate}},
  \href{http://dx.doi.org/10.1142/S0217751X18501440}{\emph{Int. J. Mod. Phys.
  A} {\bfseries 33} (2018) 1850144},
  [\href{https://arxiv.org/abs/1803.01249}{{\ttfamily 1803.01249}}].

\bibitem{Planck:2018vyg}
{\scshape Planck} collaboration, N.~Aghanim et~al., \emph{{Planck 2018 results.
  VI. Cosmological parameters}},
  \href{http://dx.doi.org/10.1051/0004-6361/201833910}{\emph{Astron.
  Astrophys.} {\bfseries 641} (2020) A6},
  [\href{https://arxiv.org/abs/1807.06209}{{\ttfamily 1807.06209}}].

\bibitem{Weinberg:2013kea}
S.~Weinberg, \emph{{Goldstone Bosons as Fractional Cosmic Neutrinos}},
  \href{http://dx.doi.org/10.1103/PhysRevLett.110.241301}{\emph{Phys. Rev.
  Lett.} {\bfseries 110} (2013) 241301},
  [\href{https://arxiv.org/abs/1305.1971}{{\ttfamily 1305.1971}}].

\bibitem{Planck:2015fie}
{\scshape Planck} collaboration, P.~A.~R. Ade et~al., \emph{{Planck 2015
  results. XIII. Cosmological parameters}},
  \href{http://dx.doi.org/10.1051/0004-6361/201525830}{\emph{Astron.
  Astrophys.} {\bfseries 594} (2016) A13},
  [\href{https://arxiv.org/abs/1502.01589}{{\ttfamily 1502.01589}}].

\bibitem{AMS:2021nhj}
{\scshape AMS} collaboration, M.~Aguilar et~al., \emph{{The Alpha Magnetic
  Spectrometer (AMS) on the international space station: Part II \textemdash{}
  Results from the first seven years}},
  \href{http://dx.doi.org/10.1016/j.physrep.2020.09.003}{\emph{Phys. Rept.}
  {\bfseries 894} (2021) 1--116}.

\bibitem{Giesen:2015ufa}
G.~Giesen, M.~Boudaud, Y.~G\'enolini, V.~Poulin, M.~Cirelli, P.~Salati et~al.,
  \emph{{AMS-02 antiprotons, at last! Secondary astrophysical component and
  immediate implications for Dark Matter}},
  \href{http://dx.doi.org/10.1088/1475-7516/2015/9/023}{\emph{JCAP} {\bfseries
  09} (2015) 023}, [\href{https://arxiv.org/abs/1504.04276}{{\ttfamily
  1504.04276}}].

\bibitem{Bergstrom:2013jra}
L.~Bergstrom, T.~Bringmann, I.~Cholis, D.~Hooper and C.~Weniger, \emph{{New
  Limits on Dark Matter Annihilation from AMS Cosmic Ray Positron Data}},
  \href{http://dx.doi.org/10.1103/PhysRevLett.111.171101}{\emph{Phys. Rev.
  Lett.} {\bfseries 111} (2013) 171101},
  [\href{https://arxiv.org/abs/1306.3983}{{\ttfamily 1306.3983}}].

\bibitem{HESS:2022ygk}
{\scshape H.E.S.S.} collaboration, H.~Abdalla et~al., \emph{{Search for Dark
  Matter Annihilation Signals in the H.E.S.S. Inner Galaxy Survey}},
  \href{http://dx.doi.org/10.1103/PhysRevLett.129.111101}{\emph{Phys. Rev.
  Lett.} {\bfseries 129} (2022) 111101},
  [\href{https://arxiv.org/abs/2207.10471}{{\ttfamily 2207.10471}}].

\bibitem{McDaniel:2023bju}
A.~McDaniel, M.~Ajello, C.~M. Karwin, M.~Di~Mauro, A.~Drlica-Wagner and M.~A.
  S\'anchez-Conde, \emph{{Legacy analysis of dark matter annihilation from the
  Milky~Way dwarf spheroidal galaxies with 14~years of Fermi-LAT data}},
  \href{http://dx.doi.org/10.1103/PhysRevD.109.063024}{\emph{Phys. Rev. D}
  {\bfseries 109} (2024) 063024},
  [\href{https://arxiv.org/abs/2311.04982}{{\ttfamily 2311.04982}}].

\bibitem{HESS:2013rld}
{\scshape H.E.S.S.} collaboration, A.~Abramowski et~al., \emph{{Search for
  Photon-Linelike Signatures from Dark Matter Annihilations with H.E.S.S.}},
  \href{http://dx.doi.org/10.1103/PhysRevLett.110.041301}{\emph{Phys. Rev.
  Lett.} {\bfseries 110} (2013) 041301},
  [\href{https://arxiv.org/abs/1301.1173}{{\ttfamily 1301.1173}}].

\bibitem{Albert:2014hwa}
{\scshape Fermi-LAT} collaboration, A.~Albert, G.~A. Gomez-Vargas, M.~Grefe,
  C.~Munoz, C.~Weniger, E.~D. Bloom et~al., \emph{{Search for 100 MeV to 10 GeV
  $\gamma$-ray lines in the Fermi-LAT data and implications for gravitino dark
  matter in $\mu\nu$SSM}},
  \href{http://dx.doi.org/10.1088/1475-7516/2014/10/023}{\emph{JCAP} {\bfseries
  10} (2014) 023}, [\href{https://arxiv.org/abs/1406.3430}{{\ttfamily
  1406.3430}}].

\end{thebibliography}\endgroup

\end{document}